\documentclass[a4paper,fleqn,usenatbib]{mnras}

\usepackage{mathptmx}

\usepackage[T1]{fontenc}
\usepackage{ae,aecompl}

\usepackage{float}
\usepackage{graphicx}	
\usepackage{mathrsfs,amsmath}	
\usepackage{amssymb}	
\usepackage{algorithm}
\usepackage{algpseudocode} 
\usepackage{blindtext}
\usepackage{caption}
\usepackage{subcaption}
\usepackage{adjustbox}
\usepackage{array}
\usepackage{booktabs}
\usepackage{enumitem}
\graphicspath{{./ImagesForPaper/}}
\title[Performance Assessment of Pulsar Search Pipelines]{A Framework for Assessing the Performance of Pulsar Search Pipelines}

\author[E. van Heerden et al.]{
E. van Heerden,$^{1}$\thanks{E-mail: elmarie.vanheerden@eng.ox.ac.uk}
A. Karastergiou,$^{2,3,4}$
S. J. Roberts$^{1}$
\\
$^{1}$Information Engineering, University of Oxford, Parks Road, Oxford OX1 3PJ, UK\\
$^{2}$Astrophysics, University of Oxford, Denys Wilkinson Building, Keble Road, Oxford OX1 3RH, UK\\
$^{3}$Physics Department, University of the Western Cape, Cape Town 7535, South Africa \\
$^{4}$Department of Physics and Electronics, Rhodes University, PO Box 94, Grahamstown 6140, South Africa
}

\date{Accepted 2016 November 22. Received 2016 November 22; in original form 2016 June 14}

\pubyear{2016}

\begin{document}
\label{firstpage}
\pagerange{\pageref{firstpage}--\pageref{lastpage}}
\maketitle

\begin{abstract}
In this paper, we present a framework for assessing the effect of non-stationary Gaussian noise and radio frequency interference (RFI) on the signal to noise ratio, the number of false positives detected per true positive and the sensitivity of standard pulsar search pipelines. The results highlight the necessity to develop algorithms that are able to identify and remove non-stationary variations from the data before RFI excision and searching is performed in order to limit false positive detections. The results also show that the spectrum whitening algorithms currently employed, severely affect the efficiency of pulsar search pipelines by reducing their sensitivity to long period pulsars.
\end{abstract}

\begin{keywords}
methods: analytical -- methods: data analysis -- methods: statistical -- (stars:) pulsars: general.
\end{keywords}



\section{INTRODUCTION}
\label{sec:Introduciton}
Pulsars provide a wealth of information about neutron star physics, the interstellar medium and stellar evolution \citep{b4}. Furthermore, their clock-like properties allow for sensitive measurements of their orbital dynamics which are used to constrain the equation of state of ultra-dense matter \citep{b3,b2}, probe the physics of binary evolution and test the predictions of General Relativity \citep{b1}. The continued discovery of new pulsars through pulsar surveys is paramount if we are to improve our understanding of the radio pulsar population as well as expand research in the aforementioned areas. Consequently, pulsar surveys remain a driving force in the field of astrophysics. 

Pulsar research has in the past been driven by a number of large-scale surveys carried out with various radio telescopes. Surveys of the Galactic plane \citep{b5,b21}, supernova remnants \citep{b22}, globular clusters \citep{b23} and all-sky surveys \citep{b24,b25} have led to the discovery of more than 2200 pulsars.

Pulsar population synthesis models \citep{b42}, based on pulsar surveys and the known pulsar population, are used to predict the number of pulsars expected to be discovered in future pulsar surveys \citep{b60,b61}. These techniques are also used to estimate the number of potentially detectable (i.e. those that are beaming towards us as well as being luminous enough) normal pulsars and millisecond pulsars (MSPs) in the Galaxy.

The number of pulsars actually discovered in recent surveys \citep{b43,b31} has fallen well short of the number predicted by the aforementioned estimation techniques. It was predicted that the Arecibo PALFA Precursor survey \citep{b43} should have detected $490_{-115}^{+160}$ normal pulsars and $12_{-5}^{+70}$ millisecond pulsars (MSPs) by the beginning of 2014, but managed to detect only $283$ normal pulsars and 31 MSPs. The full PALFA survey, when complete, is expected to have detected $1000_{-230}^{+330}$ normal pulsars and $30_{-20}^{+200}$ MSPs. However, close to completion it has only managed to detect $\sim 443$ normal pulsar and $\sim 40$ MSPs respectively \citep{b31}. It is worth noting that the largest discrepancy between predictions and detections is for normal pulsars, i.e. pulsars with long periods. Furthermore, it is estimated that there are between 82,000 to 143,000 detectable normal pulsars and 9,000 to 100,000 detectable MSPs in the Galactic disk alone \citep{b60,b43}, yet to date we have only discovered some 2200 pulsars \citep{b44}. The deficiencies in pulsar detections have been attributed to RFI and scintillation \citep{b42}, both these effects are not addressed in current population synthesis models \citep{b48}. 

Electromagnetic radiation with frequencies between circa 10 kHz and 100 GHz is referred to as radio frequency (RF). Radio frequency interference (RFI) in the context of a pulsar survey is any signal or disturbance emitted from a man-made source either extra-terrestrial or terrestrial that corrupts the measurements of data obtained. The spatial and temporal variability of RFI make it difficult to identify and to mitigate. If RFI is not dealt with then spurious trends may occur in the data collected, thereby decreasing the signal to noise ratio (SNR) and making it more difficult, or impossible, to detect new pulsars.

All-sky pulsar surveys, such as the Arecibo PALFA survey, are more often than not conducted in the L-band (the 1 to 2 GHz range of the radio spectrum), more specifically the frequency range 1.2 GHz to 1.6 GHz \citep{b31,b45}. The frequency range 1.2 GHz to 1.6 GHz happens to overlap with frequencies that have been earmarked for other applications such as satellite navigation, telecommunication, aircraft surveillance, amateur radio and digital audio broadcasting \citep{b50}. Most of the aforementioned RFI sources severely decrease the sensitivity of surveys conducted in the L-band.

\begin{figure} 
      \centering
      \includegraphics[width=\columnwidth]{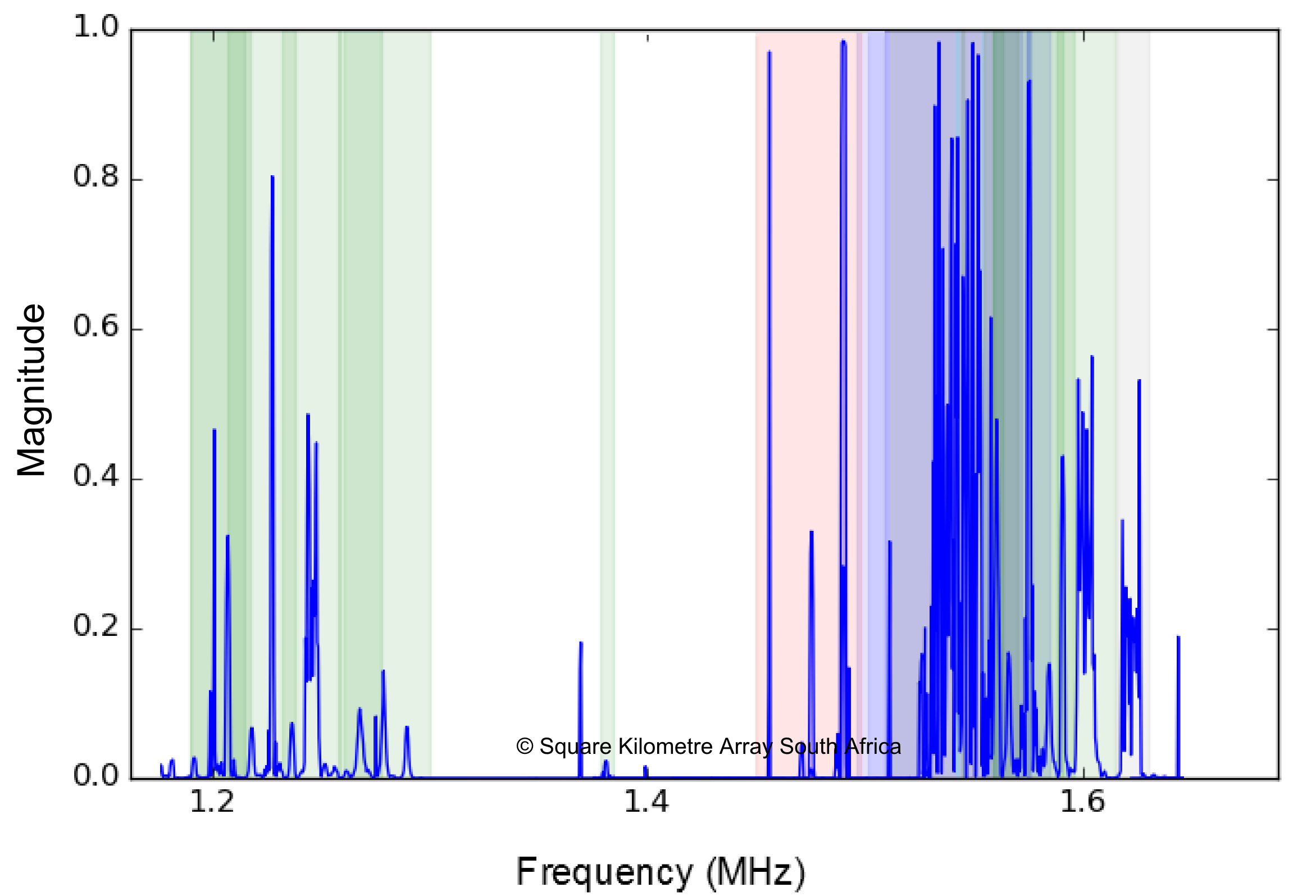}
      \caption{Typical spectrum occupancy in the L-band. \textit{Source: Square Kilometre Array South Africa}}
      \label{fig:RFI}
\end{figure}

Spectrum occupancy in the L-band, depicted in Figure~\ref{fig:RFI}, is dominated by RFI mainly from satellites. The colours in  Figure~\ref{fig:RFI} represent interference from different satellites: red - Afristar, yellow - Thuraya, blue - Inmarsat, cyan - Satellite Radio, grey - IRIDIUM, green - \{Galileo, Beidou, GPS,  GLONASS\} and grey - \{Fengun, Meteosat\}.

In a recent study by \cite{b31}, synthetic pulsars with various periods and pulse widths were injected into actual PALFA survey data with the aim to assess the effect of RFI and red noise\footnote{Red noise is a type of signal noise with a power spectral density inversely proportional to $f^2$, which means it has more energy at lower frequencies.} on the survey sensitivity. The study found that there is a significant degradation in sensitivity of between $10$~$\%$ and a factor of 2 for pulsars with spin periods between 0.1~s and 2~s and dispersion measure (DM) $>$ 150 pc cm$^{-3}$ due to red noise induced by RFI, receiver gain fluctuations and opacity variations of the atmosphere. Additionally, a population synthesis analysis based on the empirical survey sensitivity found that  $35 \pm 3$~$\%$ of pulsars, with predominantly long periods, are missed compared to expectations which are based on the theoretical sensitivity curves as derived from the radiometer equation. With these results the authors conclude that the reduced sensitivity to long-period pulsars is mainly attributed to red noise. All the results were obtained despite applying a red noise suppression algorithm. 

In this paper we show, supplementary to the \cite{b31} study, that frequency dependent noise such as red noise indeed reduces the SNR of long-period pulsars and increases the number of false detections \citep{lyon2016fifty}. Moreover, we offer an explanation as to how the red noise suppression technique in the \cite{b31} paper actually contributed to the reduced sensitivity of long-period pulsars by explaining what the algorithm does and quantifying the loss of signal to noise ratio when the algorithm is applied.

It is evident, from the number of pulsars missed, that RFI and frequency dependent noise greatly affect the sensitivity of radio telescopes to normal pulsars, i.e. pulsars with long periods. Therefore, the aims of this study are:
\begin{enumerate}[label=(\alph*),wide = 0pt, labelwidth = 1.3333em, labelsep = 0.6333em, leftmargin = \dimexpr\labelwidth + \labelsep\relax ]
\item to quantify the effect that non-stationary Gaussian noise and RFI has on the performance of pulsar search pipelines;
\item to examine the effectiveness of the current spectrum whitening methods available in pulsar search software suites;
\item to determine if detrending the data with a moving average filter before searching for pulsars is effective;
\item to examine the effectiveness of the current RFI detection and mitigation methods available in pulsar search software suites.
\item to investigate the reduction in sensitivity as a function of both the correlation length of the non-stationary noise and the pulse period. 
\end{enumerate}

We start in \S~\ref{sec:PipelineForNormalPulsars} by describing the building blocks of a typical pulsar search pipeline for normal pulsars. There exist various software implementations of this pipeline. The ones used in this analysis are introduced in \S~\ref{sec:SoftwareSuites}. Although these software packages differ in a number of ways, the way in which they deal with frequency dependent noise is of particular interest for this study. Therefore, in section \S~\ref{sec:SpectrumWhitening} the different spectrum whitening algorithms available in the software packages considered are mathematically described. The method used for generating the synthetic filterbank files with non-stationary Gaussian noise are detailed in \S~\ref{sec:NonStationaryGaussianNoise} and the RFI added to some of the files can be found in \S~\ref{sec:RFI}. In \S ~\ref{sec:ObservationParam} to \S~\ref{sec:Pipelines} we introduce the experimental framework we used for generating and processing filterbank files containing different noise processes.

The aim of this framework is to assess the ability of different pulsar search pipelines to detect pulsars embedded in non-stationary Gaussian noise amidst RFI. It is worth noting that this study differs from the \cite{b31} study in that the aim is to quantify the sensitivity of different pulsar search pipelines as a function of noise correlation length and pulsar spin period, whereas the latter aimed at quantifying the Arecibo PALFA survey's sensitivity as a function of DM and pulsar spin period. Finally, the results, discussion and conclusions can be found in \S~\ref{sec:Results} to \S~\ref{sec:Conclusion}.

\section{Search process for normal pulsars}
\label{sec:SearchForNormalPulsars}
In order to determine how phenomena like RFI and non-stationary noise can hinder pulsar detections it is necessary to first understand the nature of the acquired data and the functionality of each of the building blocks that form part of a pulsar search pipeline. Therefore, a detailed description of a typical pulsar search pipeline is presented in this section. Two existing software implementations of the pulsar search pipeline and a detailed algorithmic description of the spectrum whitening techniques available in each of them are also presented here. Different configurations of these software suites are used in a subsequent section to process pulsar data where the results will be used to assess their abilities to deal with non-stationary noise and RFI. 

\subsection{Search data}
\label{sec:SearchData}
The data that are searched for pulsars are time series of total power per frequency channel typically referred to as filterbank data. The number of frequency channels, the temporal resolution and the dynamic range (i.e. 1-bit, 8-bit or 16-bit) of the data are unique to each survey.

Files that contain filterbank data are currently processed off-line; however, with the increase in scope and sensitivity of future surveys it will become infeasible to store the raw data for off-line processing due to capacity and input/output constraints. Hence, the need for a paradigm shift from off-line to real-time processing of survey data.

Real-time processing entails block-wise dedispersion for the purposes of rapid reporting of Fast Radio Burst (FRB) detections, which constitute a byproduct of pulsar searches. The time duration of each block depends on the dispersion measure search and the observing frequency, and is likely to be of order a few seconds for typical searches. RFI cleaning must happen prior to dedispersion. Thereafter, all frequency information is lost and likewise the opportunity to detect and mitigate RFI in the time-frequency plane. The relevant timescale on which any RFI excision technique needs to operate is therefore more likely to be related to the FRB detection buffer size, rather than the full integration required for periodicity searches.


\subsection{Pipeline for a standard pulsar search}
\label{sec:PipelineForNormalPulsars}
Detecting radio pulses produced by pulsars is an intrinsically difficult task due to their narrow duty cycles, low signal strengths, dispersion effects and the presence of non-Gaussian noise.

Numerous techniques have been developed to overcome some of the difficulties highlighted above. These techniques are combined to form the standard pulsar search pipeline.  

\begin{figure*}
      \centering
      \includegraphics[width=\textwidth]{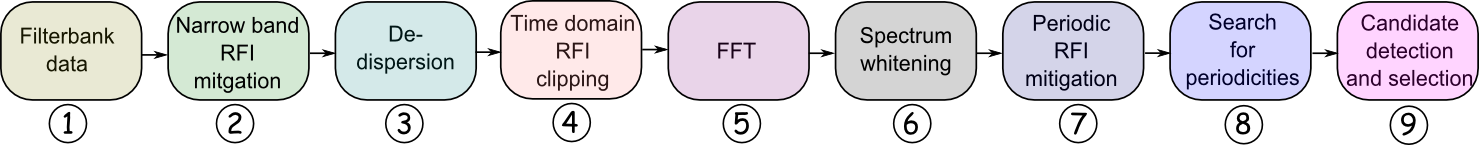}
      \caption{Schematic illustration of a typical pulsar search pipeline, see text for details.}
      \label{fig:SearchPipeline}
\end{figure*}

The typical pipeline used for the processing of filterbank files consists of nine stages as depicted in  Figure~\ref{fig:SearchPipeline}. The pipeline starts with noisy filterbank files (Figure~\ref{fig:SearchPipeline}.1) that may or may not contain a pulsar. 

Each filterbank file is examined for narrowband RFI signals, which are excised by replacing the affected samples with constant values chosen to match the median bandpass (Figure~\ref{fig:SearchPipeline}.2) \citep{b31}. 

The corrected filterbank files are then dedispersed for a number of trial dispersion measure (DM) values to compensate for the dispersion induced by the interstellar medium (Figure~\ref{fig:SearchPipeline}.3) \citep{b4}. The zero-DM time series is used to identify and mitigate broad band RFI (Figure~\ref{fig:SearchPipeline}.4) that went undetected by the narrow band RFI excision process. After mitigating broad band RFI, the Fourier transform of each dedispersed time series is computed (Figure~\ref{fig:SearchPipeline}.5). 

The power spectrum is whitened (Figure~\ref{fig:SearchPipeline}.6) so that the response is as uniform as possible, i.e., mitigating frequency dependent noise, subtracting a running median and normalising the local root mean square (rms) of the power spectrum such that it has a zero mean and unit rms. A whitened power spectrum is preferred, because estimating the significance level of any signal present is relatively easy. Different techniques have been implemented to whiten the spectrum and these will be described in more detail in section \ref{sec:SpectrumWhitening}.

The next stage of the pipeline concerns identification of periodic RFI. Known periodic signals which are present all or most of the time, such as power lines carrying alternating current and communication systems such as airport radar systems are flagged with their harmonics and their bandwidths determined. These interferences are mitigated by creating a spectral mask (see Figure~\ref{fig:SearchPipeline}.7). This mask consists of a list of all the Fourier bins affected and which should be ignored in all subsequent processing.

Radio pulses from pulsars generally have narrow duty cycles which, in the Fourier domain, results in the power to be distributed between the fundamental frequency and a number of harmonics \citep{b29}. Therefore, to take full advantage of the power contained in the harmonics the whitened spectrum is harmonically summed by adding the higher harmonics to the fundamentals. The original power spectra as well as the composite spectra formed by summing 2, 4, 8 and 16 harmonics \citep{b40} are each searched for periodicities (Figure~\ref{fig:SearchPipeline}.8) \citep{b25}. The best candidates from each trial DM are saved.

After all the time series have been processed, a list of pulsar candidates is compiled. This list is pruned by post-processing procedures (Figure~\ref{fig:SearchPipeline}.9) ranging from sifting and folding to sophisticated machine learning candidate selection \citep{b31}. The most promising pulsar candidates are saved for future observation and follow-up \citep{b25}.

\subsection{Pulsar search software} 
\label{sec:SoftwareSuites}
The pulsar search pipeline described above is available in a number of pulsar search software packages: \texttt{SIGPROC} developed by Lorimer  \citep{b30}, \texttt{PRESTO} developed by Ransom \citep{b39}, \texttt{PEASOUP} developed by Barr \citep{Peasoup} and \texttt{PULSARHUNTER} developed by Keith \citep{KeithThesis}. The two most frequently used packages are \texttt{SIGPROC} and \texttt{PRESTO}, both of which are freely available and well tested. Together they have been responsible for the discovery of most of the pulsars known today. We refer the interested reader to \cite{cordes2006arecibo}, \cite{rane2016search}, \cite{stovall2014green} and \cite{b31} for a comprehensive discussion of how these two pipelines are typically used in real pulsar surveys.

There are three main differences between \texttt{SIGPROC} and \texttt{PRESTO}. Firstly, the manner in which they search for pulsars that orbit a companion, namely binary pulsars. Radio pulses from binary pulsars typically exhibit Doppler shifts in their rotational period, caused by the acceleration of the pulsar around its companion. To be efficient, these shifts need to be accounted for in the so-called acceleration searches. \texttt{PRESTO} performs acceleration searches in the Fourier domain \citep{b63}. \texttt{SIGPROC} on the other hand does time-domain re-sampling to carry out acceleration searches. Secondly, \texttt{SIGPROC} looks for harmonically related signals in the amplitude spectrum, whereas \texttt{PRESTO} uses the power spectrum to identify possible pulsar candidates \citep{b4}. Lastly, \texttt{SIGPROC} uses SNR as a metric to identify peaks in the normalised power spectrum whereas \texttt{PRESTO} uses the Gaussian significance (adjusted for the number of trials searched) of the peaks as a metric under a white noise assumption. Hereinafter, the terms SNR and Gaussian significance shall be collectively referred to as \textit{detection significance}.

\subsection{Spectrum whitening}
\label{sec:SpectrumWhitening}
A stochastic process is considered white if and only if it is stationary and independent at all points. As a consequence, the power spectral density of a white process is uniformly distributed across the whole available frequency range. A non-white process is instead characterised by a given distribution of the power per unit frequency along the available frequency bandwidth \citep{b31}. A whitening operation on any non-white process entails forcing said process to adhere to the conditions described above for a white process. 

In the case of pulsar searching, well-behaved white noise is sought after, because it simplifies any attempt at estimating the significance levels of any signal present in the data and consequently makes detection easier. Hence, it is standard practice to whiten the power spectral density by suppressing frequency-dependant noise, in particular red noise, so that the response to noise is as uniform as possible. 

The spectrum whitening techniques implemented in \texttt{SIGPROC} and \texttt{PRESTO} are similar in that they aim to normalise the spectrum. However, the way in which these techniques algorithmically operate in normalising the spectrum is quite different. The different whitening options available in \texttt{SIGPROC} and \texttt{PRESTO} are mathematically described in the two subsequent sections.

\subsubsection{Spectrum whitening in SIGPROC}
\begin{figure}
      \centering
      \includegraphics[width=\columnwidth]{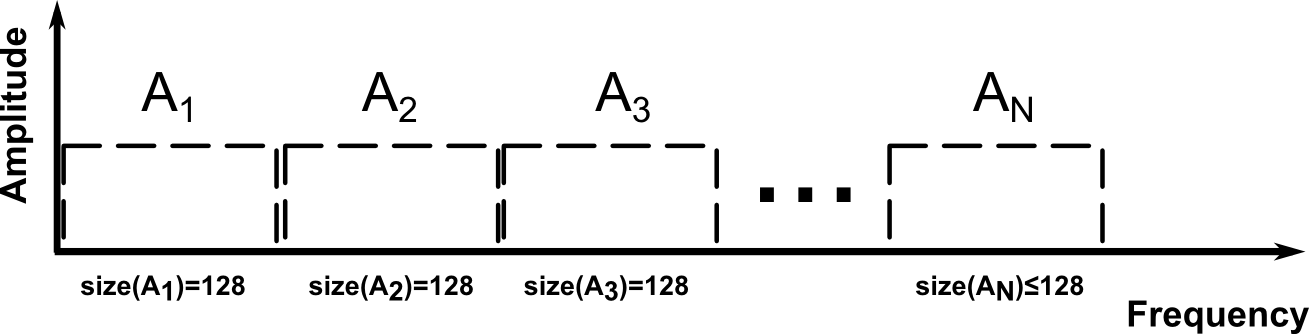}
      \caption{Amplitude spectrum partitioning for the whitening algorithm implemented in \texttt{SIGPROC}, see text for details.}
      \label{fig:SIGPROCwhitening}
\end{figure}

In \texttt{SIGPROC} there are three spectrum whitening options for the \texttt{SEEK} function. In all three approaches the spectrum is divided into blocks of \mbox{max}$\{128, (\text{number of spectral data points}/400000)\}$ Fourier bins. The simulation parameters of this analysis resulted in the number of Fourier bins per block to be consistently 128 and thus for all subsequent explanations we assume that the spectrum is divided into blocks of 128 Fourier bins as depicted in Figure~\ref{fig:SIGPROCwhitening}.

The algorithms implemented for the three options in \texttt{SIGPROC} are:
\begin{itemize}[wide, labelwidth=!, labelindent=0pt]
\item Option 1:\\
The default whitening algorithm executed when the function \texttt{SEEK} is called computes the mean and corrected sample standard deviation, $s = \sqrt {{\raise0.7ex\hbox{$1$} \!\mathord{\left/ {\vphantom {1 {(N - 1)}}}\right.\kern-\nulldelimiterspace}\!\lower0.7ex\hbox{${(N - 1)}$}}\sum\nolimits_{i = 1}^N {{{({x_i} - \bar x)}^2}} }$, for each of the blocks A$_{1}$,...,A$_\text{N}$ in Figure~\ref{fig:SIGPROCwhitening}. The mean is subtracted from each Fourier bin in the block, whereafter the bins are scaled by the corrected sample standard deviation of that particular block. The algorithmic steps are detailed in Algorithm~\ref{algor:default}. 
\begin{algorithm}
\caption{SIGPROC: \texttt{default}}
\label{algor:default}
\begin{algorithmic}[1]
	\For {$i=1,...,N$}
		\State $\mu_{i} = $ mean(A$_{i}$)
		\State s$_{i} = $ corrected standard deviation(A$_{i}$)
		\State $\text{A}_{i}^{\text{new}}=(\text{A}_{i}^{\text{old}}-\mu_{i})/\text{s}_{i}$
		\EndFor
\end{algorithmic}
\end{algorithm}
\item Option 2: \\
The whitening algorithm executed when the function \texttt{SEEK} is called with the flag \texttt{-submn} is identical to Algorithm~\ref{algor:default} except for one difference. The blocks A$_{1}$,...,A$_\text{N}$ in Figure~\ref{fig:SIGPROCwhitening}  are scaled with the root mean square (rms) of that particular block instead of the corrected sample standard deviation. 

\item Option 3: \\
The whitening algorithm executed when the function \texttt{SEEK} is called with the flag \texttt{-submjk} computes the mean and corrected sample standard deviation for each of the blocks A$_{1}$,...,A$_\text{N}$ in Figure~\ref{fig:SIGPROCwhitening}. Thereafter, the gradients of the mean and corrected sample standard deviation between adjacent blocks of 128 Fourier bins are computed. For each Fourier bin $j=1,...,128$ in a block the mean is subtracted and the result scaled with the corrected sample standard deviation, whereafter the mean and corrected sample standard deviation is updated with the calculated gradients for that particular block. The algorithmic steps are detailed in Algorithm~\ref{algor:submjk}.

\begin{algorithm}
\caption{SIGPROC: \texttt{submjk}}
\label{algor:submjk}
\begin{algorithmic}[1]
	\For {$i=1,...,N$}
		\State $\mu_{i} = $ mean(A$_{i}$)
		\State $\mu_{i+1} = $ mean(A$_{i+1}$)
		\State $\text{s}_{i} = \text{corrected standard deviation}(\text{A}_{i})$
		\State $\text{s}_{i+1} = \text{corrected standard deviation}(\text{A}_{i+1})$
		\State $\text{slope}_{\text{mean}_{i}}=(\mu_{i+1}-\mu_{i})/128$
		\State $\text{slope}_{\text{s}_{i}}=(\text{s}_{i+1}-\text{s}_{i})/128$
		\For {$j=1,...,128$ in A$_{i}$}
				\State $\text{A}_{i}^{\text{new}}[j]=(\text{A}_{i}^{\text{old}}[j]-\mu_{i})/s_{i}$
				\State Update: $\mu_{i}=\mu_{i}+\text{slope}_{\text{mean}_{i}}$
				\State Update: $\text{s}_{i}=\text{s}_{i}+\text{slope}_{\text{s}_{i}}$
		\EndFor
		\EndFor
\end{algorithmic}
\end{algorithm}
\end{itemize}

\subsubsection{Spectrum whitening in PRESTO}
In \texttt{PRESTO} there is only one spectrum whitening technique implemented to suppress frequency dependent noise \citep{1538-3881-124-3-1788}. The median power level is measured in blocks across Fourier bins and then multiplied by log 2 to convert the median value to an equivalent mean level assuming that the powers are distributed exponentially. Thereafter, the measured mean power values (variable P in Algorithm~\ref{algor:Presto}) are used to compute the slope between two adjacent Fourier bins which in turn is used to normalise the complex Fourier amplitudes (variable A in Algorithm~\ref{algor:Presto}).

\begin{figure}
      \centering
      \includegraphics[width=\columnwidth]{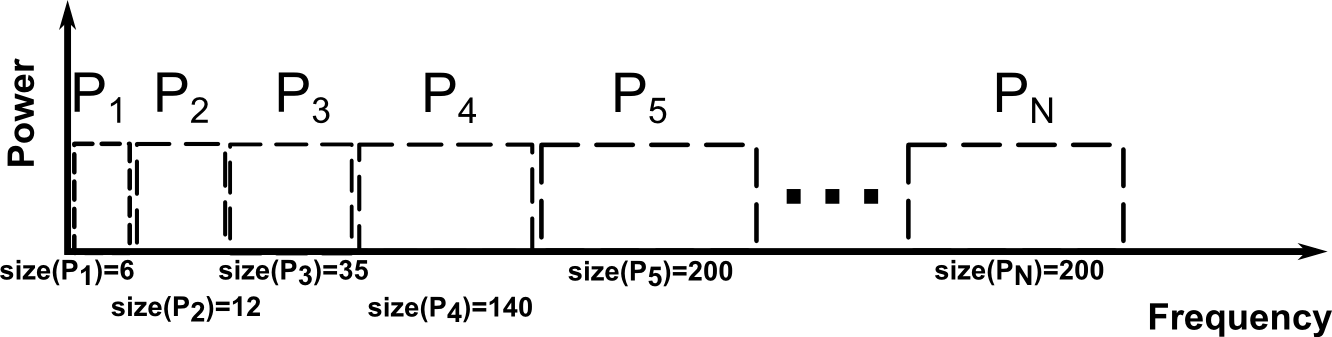}
      \caption{Power spectrum partitioning for the whitening algorithm implemented in \texttt{PRESTO}, see text for details.}
      \label{fig:PRESTOwhitening}
\end{figure}

The number of Fourier frequency bins per block starts with 6 and increases logarithmically to 200, see Figure~\ref{fig:PRESTOwhitening}. Thus, for frequencies where there is little coloured noise the number of bins used per block are 200. The algorithmic steps for the spectrum whitening technique implemented in \texttt{PRESTO} are detailed in Algorithm~\ref{algor:Presto}.
  
\begin{algorithm}
\caption{\texttt{PRESTO}: \texttt{default}}
\label{algor:Presto}
\begin{algorithmic}[1]
	\For {$i=1,...,N$}
		\State $\mu_{i} = $ median(P$_{i}$)/log 2
		\State $\mu_{i+1} = $ median(P$_{i+1}$)/log 2
		\State $\text{slope}_{i}=(\mu_{i+1}-\mu_{i})/(\text{size}(\text{P}_{i})+\text{size}(\text{P}_{i+1}))$
		\State $\text{lineoffset}=\frac{1}{2}(\text{size}(\text{P}_{i})+\text{size}(\text{P}_{i+1}))$
		\For {$j=1,...,\text{size(P}_{i})$}
				\State Update: $\text{lineval}=\mu_{i}+\text{slope}_{i} \times (\text{lineoffset}-j)$
				\State Update: $\text{scaleval}=1/\sqrt{\text{lineval}}$
				\State Update: $\textbf{Re}\text{(A)}_{i}^{\text{new}}[j]=\textbf{Re}\text{(A)}_{i}^{\text{old}}[j]\times \text{scaleval}$
				\State Update: $\textbf{Im}\text{(A)}_{i}^{\text{new}}[j]=\textbf{Im}\text{(A)}_{i}^{\text{old}}[j]\times \text{scaleval}$
		\EndFor
		\EndFor
\end{algorithmic}
\end{algorithm}
Irrespective of the red noise suppression algorithm being applied, \texttt{PRESTO} by default normalises the power spectrum using median blocks before performing harmonic summing and searching.

\section{Methodology}
\label{sec:Methodology}
Pulsar software currently available for generating synthetic pulsar data as well as the detection and timing thereof, assumes that the noise present in the signals acquired by radio telescopes is additive white Gaussian noise. This assumption ignores the fluctuating nature of the sky temperature \citep{b4,b70} as well as the effects that RFI have on the noise baselines of the data. Consequently, it is poorly understood how the aforementioned phenomena, which are clearly non-stationary, affect the ability of pulsar search pipelines to detect pulsars. Hence, the need for software to emulate these phenomena (see \S~\ref{sec:SyntheticFileGeneration}) and a framework whereby their effect on the ability of pulsar search pipelines to detect pulsars can be investigated and quantified (see \S~\ref{sec:FrameworkForPulsarAnalysis}). 

\subsection{Synthetic file generation}
\label{sec:SyntheticFileGeneration}
In section \S~\ref{sec:NonStationaryGaussianNoise} we present the low-pass filter that was inspired by a Gaussian Process \citep{b32} to generate synthetic filterbank files with non-stationary noise baselines. Additionally, in section \S~\ref{sec:RFI} we describe the choice of RFI that we injected into a subset of the synthetic filterbank files. 

\subsubsection{Non-stationary Gaussian noise}
\label{sec:NonStationaryGaussianNoise}
Filterbank files contain quantised power values computed by superimposing tens or even hundreds of single Nyquist power measurements (see Equation~\ref{eq:TotalPower}). The power measurement of a single Nyquist sample is computed from  the real and imaginary parts of the raw voltages associated with either linear or circular polarised electromagnetic waves acquired by radio telescopes. The power measurement of a single Nyquist sample is given as:
\begin{equation}
\text{Power}=X_{real}^{2}+X_{imag}^{2}+Y_{real}^{2}+Y_{imag}^{2},
\label{eq:TotalPower}
\end{equation}
where $X$ and $Y$ are either the horizontal and vertical components of linear polarisation or the left and right handed components of circular polarisation. 

The power values found in filterbank files comprise both signal and noise. The noise levels in the filterbank files are proportional to the overall system temperature which is affected by RFI, the sky temperature and the receiver temperature. During an observation various objects and RFI with different brightness levels are encountered so the duration and magnitude of the non-stationarity associated with each of these phenomena differ greatly. 

In order to generate a time series with correlated samples, i.e. a varying noise baseline, we constructed a low-pass filter (see Equation~\ref{eq:Lowpass}) and convolved it with random samples drawn from a Gaussian distribution with zero mean and unit variance ($\mathcal{N}\left( {0,1} \right)$) (see Equation~\ref{eq:unitStandardNormal}), where  $\epsilon=1\times10^{-5}$, $t$ is the sampling interval and $N$ the number of samples in the observation. Consequently, the convolution yields a vector, \textbf{w} with samples correlated over length scales defined by $\lambda$ and magnitudes proportional to $h$ (see Equation~\ref{eq:convolution}). 

\begin{equation}
 \textbf{u} := {h^2}\exp \left[ { - {{\left( {\frac{t}
 {\lambda}} \right)}^2}} \right] \forall \text{  } t \in \mathbb{R} \quad \text{such that } \quad \textbf{u}>\epsilon
 \label{eq:Lowpass}
\end{equation}
\begin{equation}
\textbf{v}: = {z_1},{z_2},...,{z_N}\sim \mathcal{N}\left( {0,1} \right)
\label{eq:unitStandardNormal}
\end{equation}
\begin{equation}
\textbf{w}=\textbf{u}\ast \textbf{v}
\label{eq:convolution}
\end{equation}

In order to generate $N$ data points which are correlated over long length scales (i.e. $\lambda >>$) requires $N$ random samples drawn from a standard normal distribution ($\mathcal{N}\left( {0,1} \right)$) to be convolved with a finely sampled low-pass filter which is compact on a large interval. Consequently, convolving two large vectors is computationally very expensive. To circumvent this challenge we generate data points with the required correlation length by convolving a fraction of the random samples drawn from a standard normal distribution with a coarsely sampled low-pass filter and then interpolating between the resultant points to produce a time-series with the desired number of points.

The vector $\textbf{w}$ generated by convolving the low-pass filter with samples drawn from a standard normal distribution is not always positive. However, for the purpose of these experiments non-negative samples are desired because the standard deviation of the baseline drift needs to be proportional to the square root of the mean. Therefore, a new vector $\textbf{g}$ is defined:
\begin{equation}
\textbf{g}=\textbf{w}-\text{min}(\textbf{w}),
\end{equation}
such that the offset is equal to zero and all the values are non-negative. 

The mean vector $\textbf{g}$ is used to generate samples for the vectors $X_{\text{real}}$, $X_{\text{imag}}$, $Y_{\text{real}}$, $Y_{\text{imag}}$:
\begin{equation}
X_{\text{real}},X_{\text{imag}},Y_{\text{real}},Y_{\text{imag}} \stackrel{i.i.d.}{\sim} \mathcal{N}  \left( \textbf{g},\sqrt{\textbf{g}} \right),
\label{eq:SamplesDrawnFromGaussian}
\end{equation}
such that the power for each sample in the synthetic filterbank file can be computed using Equation~\ref{eq:TotalPower}. 

\begin{figure*}
   			\centering
          \begin{subfigure}[b]{0.49\textwidth}
              \includegraphics[width=\textwidth]{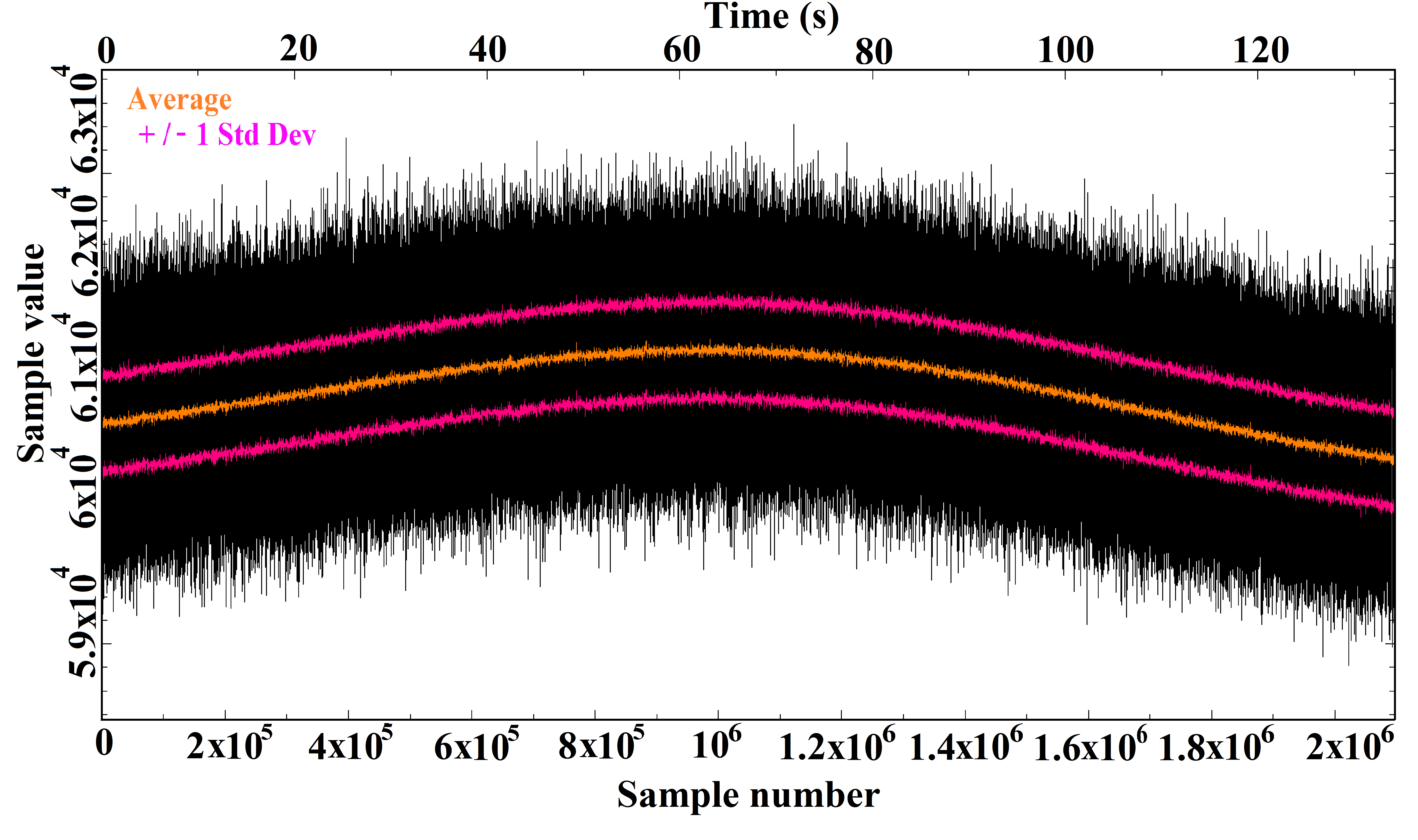}
              \caption{$\lambda = 100$ s}
              \label{fig:lambda100}
          \end{subfigure}
          ~
          \begin{subfigure}[b]{0.49\textwidth}
              \includegraphics[width=\textwidth]{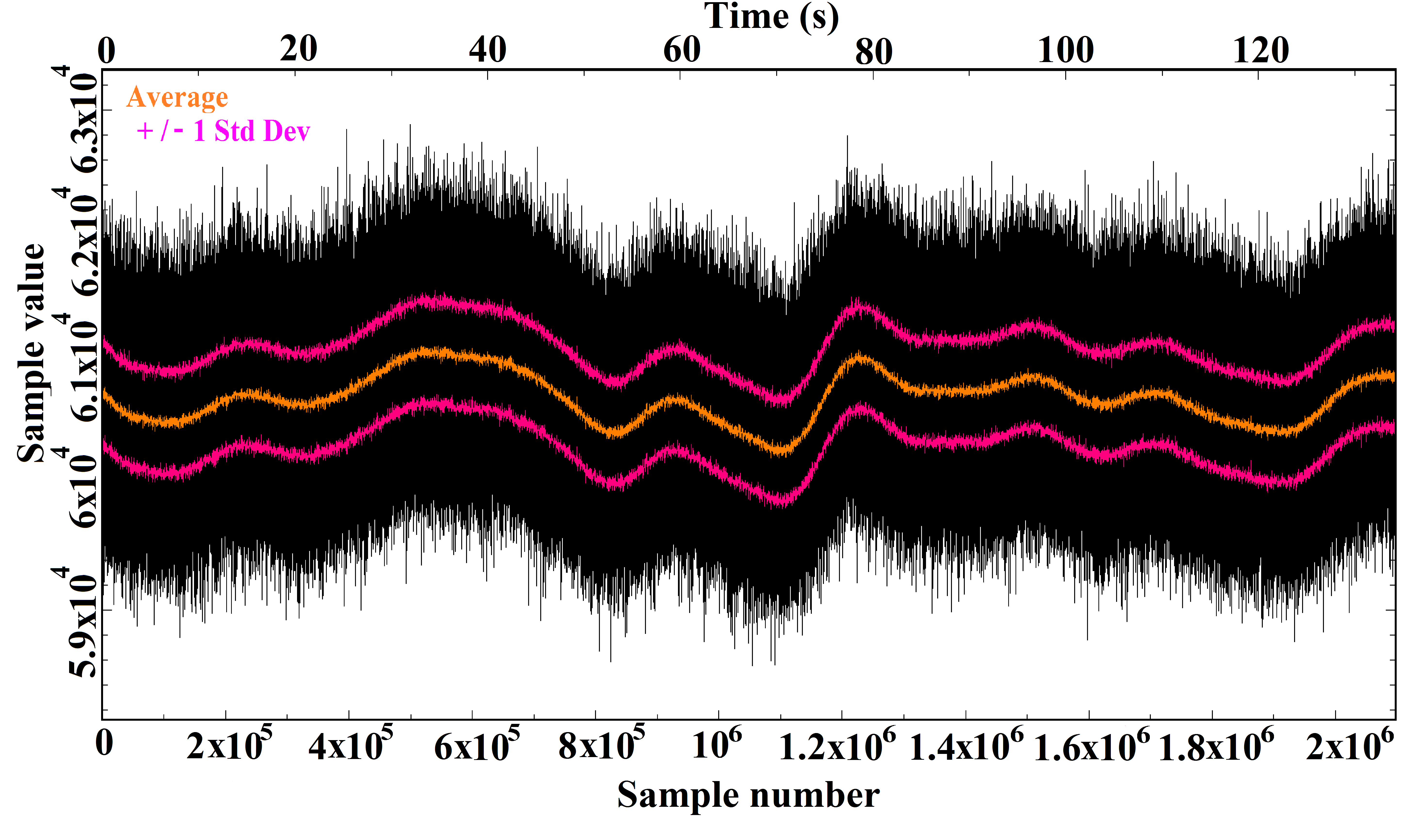}
              \caption{$\lambda = 10$ s}
              \label{fig:lambda10}
          \end{subfigure}
          
	        \begin{subfigure}[b]{0.49\textwidth}
	            \includegraphics[width=\textwidth]{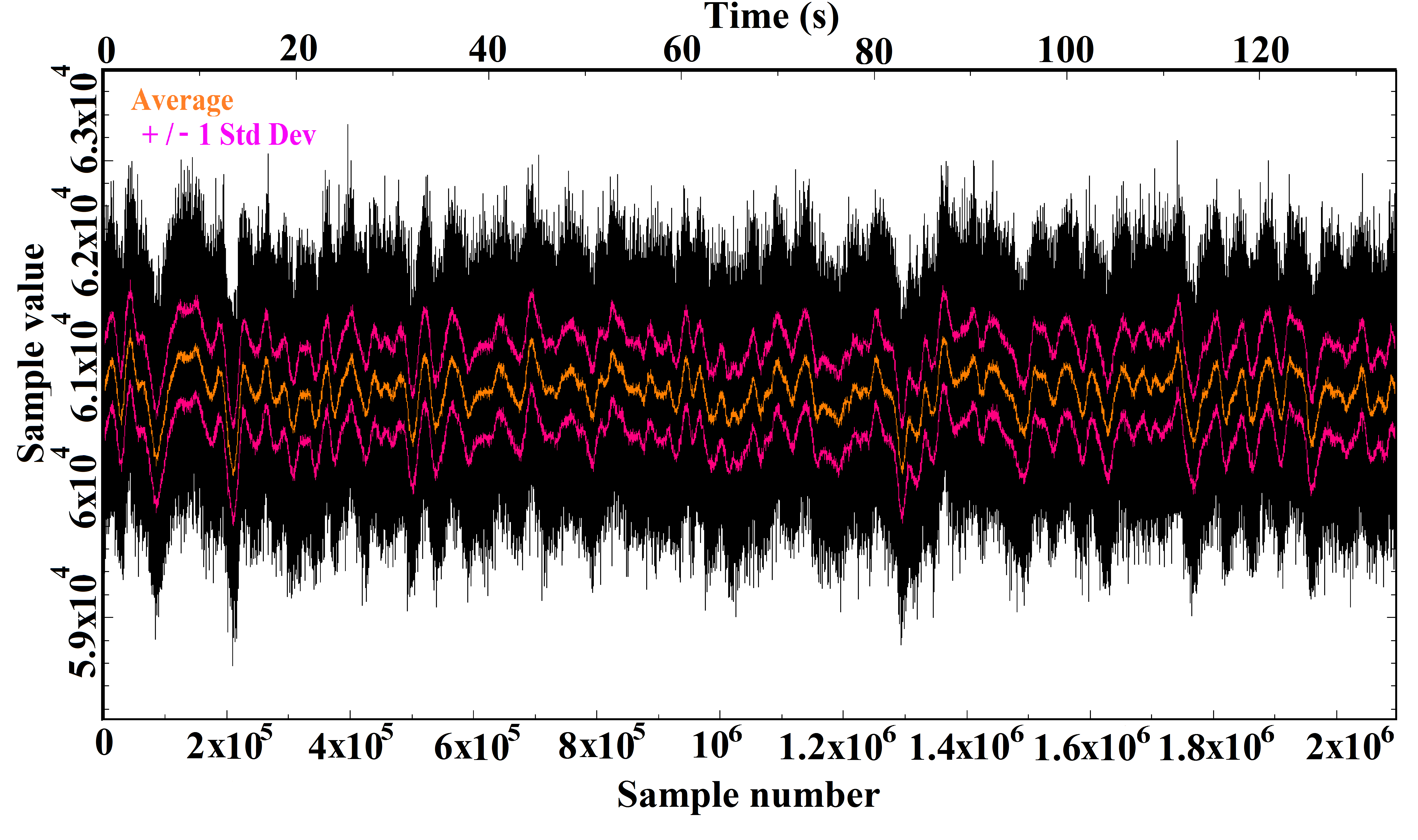}
	            \caption{$\lambda = 1$ s}
	            \label{fig:lambda1}
	        \end{subfigure}
	        ~
	        \begin{subfigure}[b]{0.49\textwidth}
	            \includegraphics[width=\textwidth]{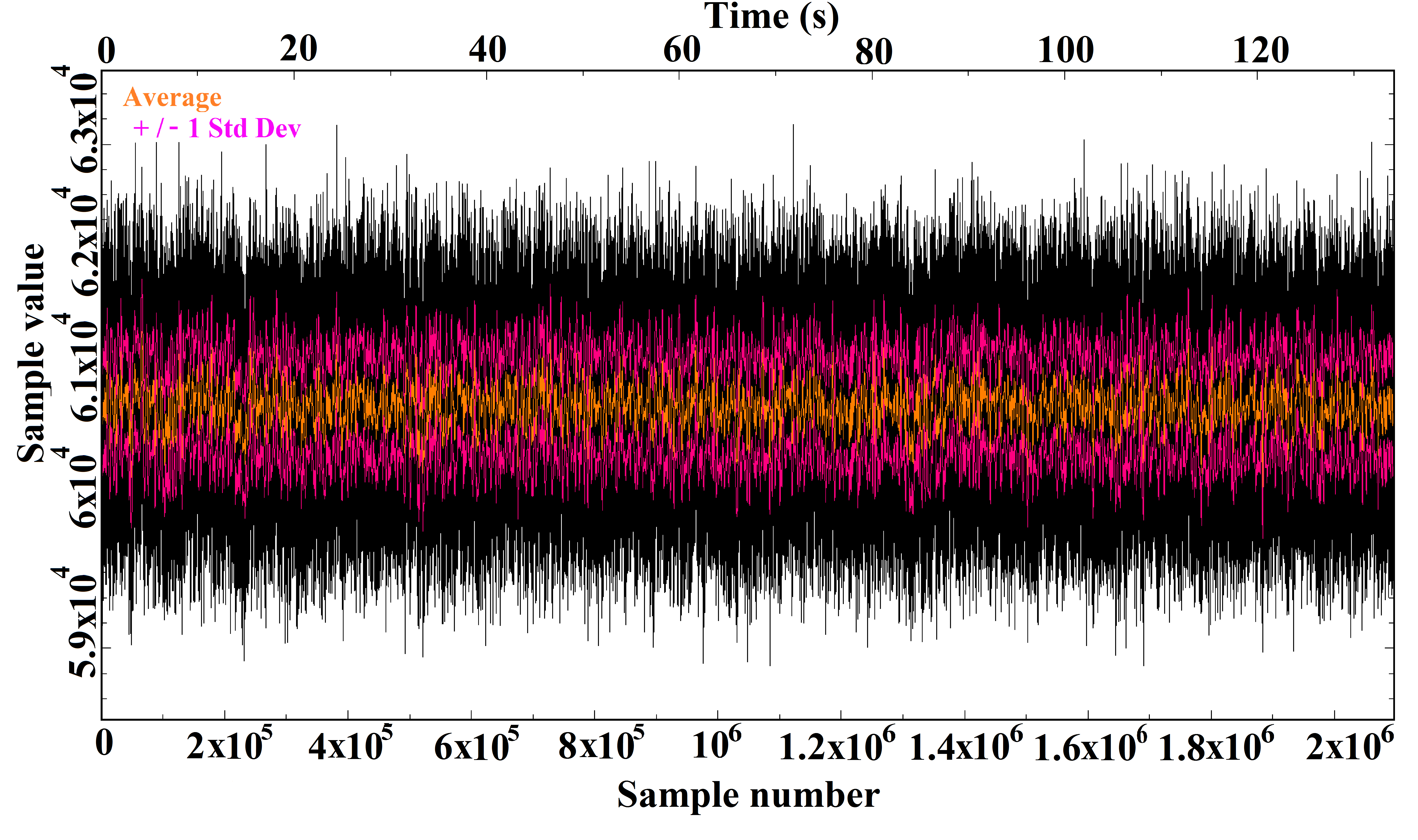}
	            \caption{$\lambda = 0.1$ s}
	            \label{fig:lambda01}
	        \end{subfigure}
	        
 	        \begin{subfigure}[b]{0.49\textwidth}
 	            \includegraphics[width=\textwidth]{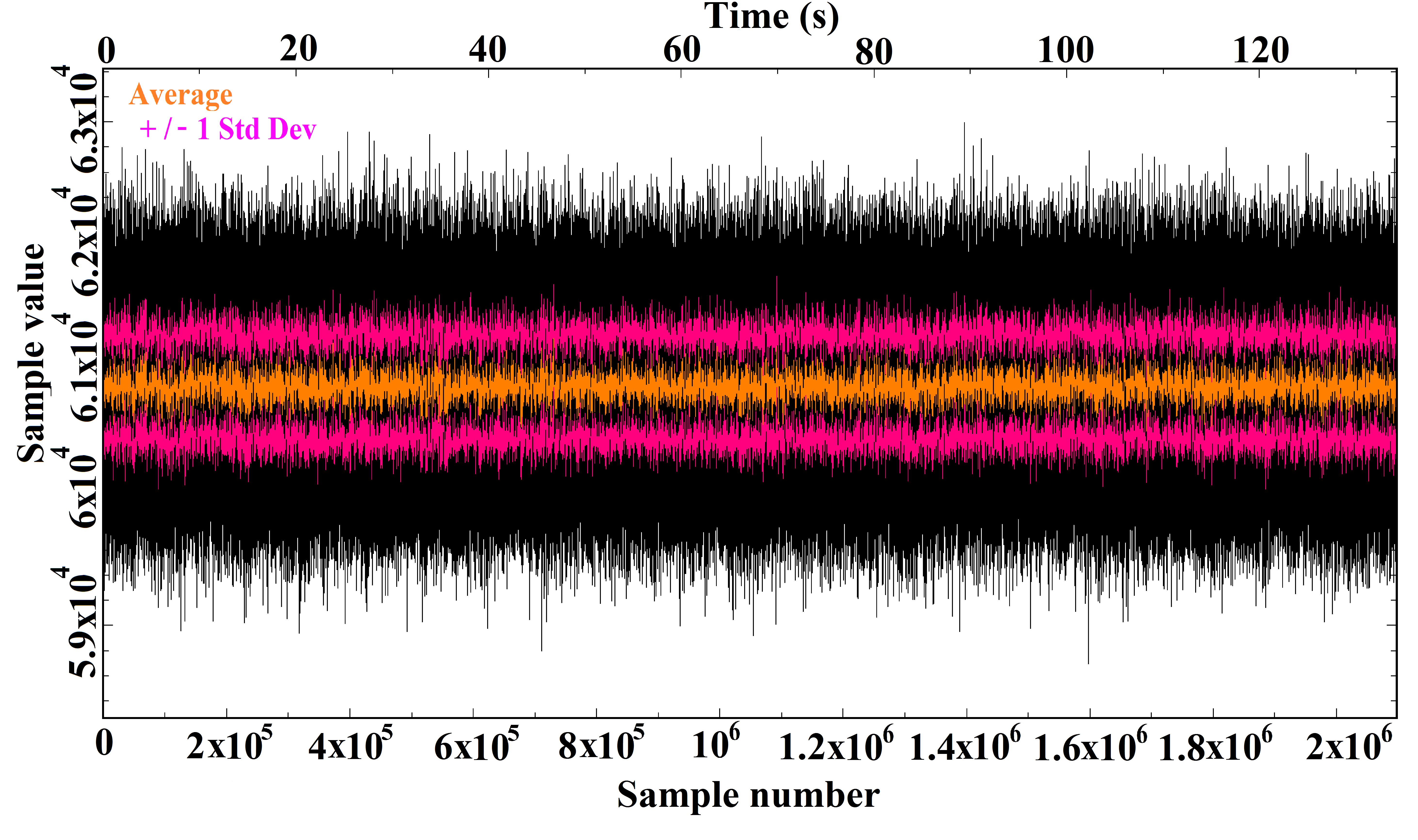}
 	            \caption{$\lambda = 0.01$ s}
 	            \label{fig:lambda001}
 	        \end{subfigure}
      \caption{ Examples of dedispersed time series that correspond to the five correlation lengths used to simulate non-stationary noise processes. Black represents the actual signal in the filterbank file, orange the mean and pink the standard deviation ($1\sigma$) of the dedispersed time series. The correlation length decreases from (a) 100~s to (e) 0.01 s.}
      \label{fig:Question6}
   \end{figure*}

A large value for the length-scale variable $\lambda$ of the low-pass filter results in a slow drifting baseline as depicted in Figure~\ref{fig:lambda100}. As the value of $\lambda$ deceases the baseline drifts becomes more capricious as depicted in Figure~\ref{fig:lambda10} to Figure~\ref{fig:lambda001}. 

\subsubsection{RFI injected}
\label{sec:RFI}

\begin{figure}
      \centering
      \includegraphics[width=\linewidth]{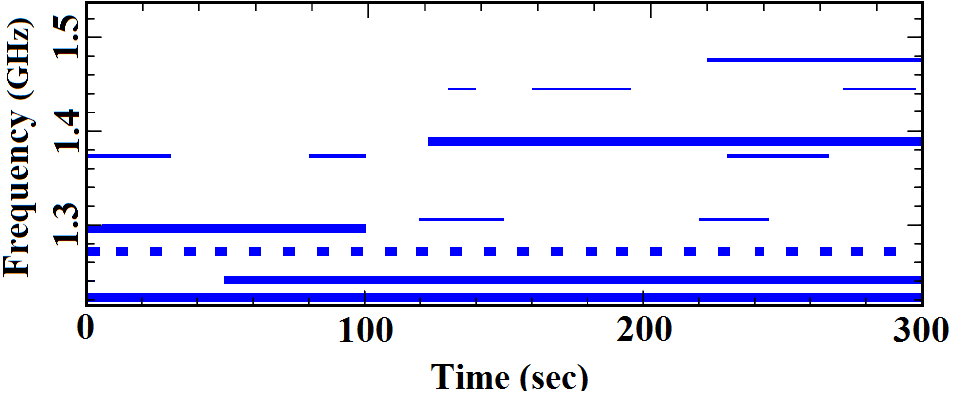}
      \caption{The RFI injected into each filterbank file.}
      \label{fig:RFIMask}
\end{figure}

For the experiments aimed at investigating the effects of RFI on the performance of pulsar search pipelines we injected the same RFI into all the filterbank files, see Figure~\ref{fig:RFIMask}. The choice of injected RFI was obtained by studying spectrum occupancy data from the KAT-7 radio telescope over several months (see Figure~\ref{fig:RFI}) and cross-correlating that with the L-band spectrum allocation as determined by the International Telecommunication Union (ITU) \citep{b50}. 
The RFI injected include:
\begin{enumerate}[label=(\alph*),wide = 0pt, labelwidth = 1.3333em, labelsep = 0.6333em, leftmargin = \dimexpr\labelwidth + \labelsep\relax ]
\item broadband periodic RFI with a period of 0.02~s and a duty cycle of 50 \%,
\item narrowband periodic RFI with a period of 12~s and a duty cycle of 25 \% affecting the frequency channels 1.266 GHz to 1.276 GHz,
\item several instances of narrowband RFI with random durations affecting the frequency channels identified from the RFI characterisation plot in Figure~\ref{fig:RFI}.
\end{enumerate}

Various routines exist in current pulsar search pipelines for excising bright RFI, but the effect of weak and unknown sources of RFI are unbeknownst to us. Therefore, the magnitudes of the various instances of injected RFI were deliberately chosen to be within one sigma of the baseline such that the effect of weak RFI on pulsar search pipelines could be investigated. The percentage of samples affected by RFI is $12$~$\%$ for each filterbank file. 
 
\subsection{Framework for Pulsar Search Pipeline Analysis}

\label{sec:FrameworkForPulsarAnalysis}
In this section a framework to generate and process non-stationary noise files with RFI is introduced. This framework allows for the understanding of how non-stationary noise processes with different correlation lengths can impede the detection of pulsars with specific periods. Additionally, it contributes to the understanding of how RFI can pass undetected through current pulsar search pipelines and the consequences of not mitigating these spurious sources of interference. 

The generation part of this framework include the synthetic observation parameters (see \S~\ref{sec:ObservationParam}), the experimental design specifics for each experiment (see \S~\ref{sec:ExperimentalSetup}) and the periods of the pulsars injected for this analysis (see \S~\ref{sec:PulsarProperties}).

The processing part of this framework includes the different configurations of \texttt{SIGPROC} and \texttt{PRESTO} that were analysed (see \S ~\ref{sec:Pipelines}). 

\subsubsection{Simulated observation parameters}
\label{sec:ObservationParam}

The observation parameters that were used for generating the synthetic filterbank files were chosen to match the Arecibo PALFA survey \citep{b31} parameters and are summarised in Table~\ref{tb:simultionParam}.

\begin{table}
\caption{Simulated observation parameters}
\label{tb:simultionParam}
\begin{center}
\begin{tabular}{ll}
Parameter & \multicolumn{1}{c}{Value} \\ 
\hline \hline
$t_{\text{obs}}$ & \multicolumn{1}{c}{$300 \text{ s}$} \\ 
$t_{\text{samp}}$ & \multicolumn{1}{c}{$64$ us} \\ 
$n_{\text{bits}}$ & \multicolumn{1}{c}{8} \\
$n_{\text{nchans}}$ & \multicolumn{1}{c}{512} \\ 
$f_{\text{low}}$ & \multicolumn{1}{c}{1214 MHz} \\ 
$f_{\text{high}}$ & \multicolumn{1}{c}{1536 MHz} \\ 
$\text{Bandwidth},\Delta f$ & \multicolumn{1}{c}{322 MHz} \\
$\text{Channel Bandwidth},\Delta f_{\text{chan}}$ & \multicolumn{1}{c}{628.91 kHz} \\
\end{tabular}
\end{center}
\end{table}

\subsubsection{Experiments}

\label{sec:ExperimentalSetup}
The experiments that we designed for analysing various configurations of \texttt{SIGPROC} and \texttt{PRESTO} are summarised in Table~\ref{tb:Experiments}. 

\begin{table*}
\caption{Summary of the experiments conducted}
\label{tb:Experiments}
\begin{center}
\begin{tabular}{llllll}
Experiment & \multicolumn{1}{c}{\# Files} & \multicolumn{1}{c}{\# Pulsars} & \multicolumn{1}{c}{Noise} & \multicolumn{1}{c}{$h$} & \multicolumn{1}{c}{$\lambda$ (s)} \\ 
\hline \hline
1. Stationary & \multicolumn{1}{c}{100} & \multicolumn{1}{c}{15} & \multicolumn{1}{c}{Stationary Gaussian}  & \multicolumn{1}{c}{-} & \multicolumn{1}{c}{-} \\ 
2. Non-stationary & \multicolumn{1}{c}{100} & \multicolumn{1}{c}{15} & \multicolumn{1}{c}{Non-stationary Gaussian} & \multicolumn{1}{c}{0.1, 0.2, 0.3, 0.4} & \multicolumn{1}{c}{0.01, 0.1, 1.0, 10.0, 100.0}  \\ 
3. Stationary +RFI & \multicolumn{1}{c}{100} & \multicolumn{1}{c}{15} & \multicolumn{1}{c}{Stationary Gaussian} & \multicolumn{1}{c}{-} & \multicolumn{1}{c}{-}  \\ 
4. Non-stationary +RFI & \multicolumn{1}{c}{100} & \multicolumn{1}{c}{15} & \multicolumn{1}{c}{Non-stationary Gaussian} & \multicolumn{1}{c}{0.1, 0.2, 0.3, 0.4} & \multicolumn{1}{c}{0.01, 0.1, 1.0, 10.0, 100.0}\\ 
5. All pulse periods per $\lambda$ & \multicolumn{1}{c}{75} & \multicolumn{1}{c}{75} & \multicolumn{1}{c}{Non-stationary Gaussian} & \multicolumn{1}{c}{0.1, 0.2, 0.3, 0.4} &  \multicolumn{1}{c}{0.01, 0.1, 1.0, 10.0, 100.0}\\
6. All pulse periods per $\lambda$ + RFI & \multicolumn{1}{c}{75} & \multicolumn{1}{c}{75} & \multicolumn{1}{c}{Non-stationary Gaussian} & \multicolumn{1}{c}{0.1, 0.2, 0.3, 0.4} & \multicolumn{1}{c}{0.01, 0.1, 1.0, 10.0, 100.0} \\
\end{tabular}%
\end{center}
\end{table*}

The design of experiments 1 to 4 is such that each one emulates a blind pulsar survey. Each experiment comprises one hundred simulated pointings with a subset of these containing a pulsar. The differences between the experiments are the type of noise processes simulated and whether RFI is present of not. 
 
Experiment 1 comprises one hundred files with stationary Gaussian noise. Fifteen of the hundred files contain an injected pulsar (see Table~\ref{tab:puslarProperties}) and the remainder are without. The results from experiments 2 to 4 will be benchmarked against the results of experiment 1 because of its idealised noise process and lack of RFI.

Experiment 2 comprises one hundred files with non-stationary Gaussian noise. Fifteen of the aforementioned files contain an injected pulsar that is unique (see Table~\ref{tab:puslarProperties}) and the remainder of the files are without a pulsar. The non-stationary Gaussian noise processes were generated according to the procedure described in \S~\ref{sec:NonStationaryGaussianNoise}. Note, every noise process is unique in that each one is defined by a different non-stationary vector, $\mathbf{g}$, and the additive Gaussian noise has zero mean and variance proportional to the square root of the non-stationary vector (see Equation~\ref{eq:SamplesDrawnFromGaussian}). The length scales, $\lambda$, for the non-stationary variability of the noise baselines range from $10^{-2}$~s to $10^{2}$~s in factors of 10, i.e. twenty files were generated per $\lambda$. Consequently, each file exhibits a unique variation because of the stochastic nature of the generation process despite having the same correlation length. 

The correlation lengths can be chosen to represent any timescale that we could consider to have an effect on the survey sensitivity to periodic pulsars, and could result from instrumental variability, to environmental effects and RFI. The power spectrum of a non-stationary noise baseline with a given correlation length will contain more power in the frequencies that correspond to that length. For this reason, we choose our length scales to sample a broad range that is relevant to the pulse periods searched, as mentioned above.  Comparing the results from experiment 2 with the results of experiment 1 enables the quantification of the effect that non-stationary Gaussian noise has on the performance of pulsar search pipelines. Moreover, experiment 2 allows for the determination of the effectiveness of the spectrum whitening techniques described in \S~\ref{sec:SpectrumWhitening} and whether or not detrending the data with a moving average filter before searching for pulsars is effective.

Experiment 3 is identical to experiment 1 except for the addition of RFI (see \S~\ref{sec:RFI}). The experimental design of experiment 3 serves to investigate the ramifications when weak RFI (see \S~\ref{sec:RFIDetectionAndMitigation}) passes undetected through a pulsar search pipeline. Furthermore, it serves to investigate the efficacy of RFI detection and mitigation algorithms currently employed.

Experiment 4 is identical to experiment 2 except for the addition of RFI (see \S~\ref{sec:RFI}). Comparing the results from experiment 4 to the results of experiments 1, 2 and 3 respectively serves to quantify the combined effect that non-stationary Gaussian noise and RFI have on the performance of pulsar search pipelines and to deduce which phenomenon has the greatest impact on said performance.

Experiment 5 comprises seventy five files in total, fifteen files per correlation length $\lambda$ (see Table~\ref{tb:Experiments}). Each one of the pulsars in Table~\ref{tab:puslarProperties} was separately injected into the fifteen files with the same correlation length and this was repeated for all five values of $\lambda$. 

Experiment 6 is identical to experiment 5 except for the addition of RFI (see \S~\ref{sec:RFI}). The experimental design of experiments 5 and 6 serves to investigate the reduction in sensitivity of pulsar search pipelines as a function of both the correlation length of the non-stationary noise and the pulse period of a pulsar.

Lastly, experiments 2, 4, 5 and 6 are repeated for four different values of the magnitude parameter $h$ defined in Equation~\ref{eq:Lowpass}, namely $h=0.1$, $0.2$, $0.3$, $0.4$.

\subsubsection{Pulsar properties}
\label{sec:PulsarProperties}
The properties of the fifteen pulsars that were randomly injected into the synthetic filterbank files were taken in part from the Arecibo sensitivity study \citep{b31} and are summarised in Table~\ref{tab:puslarProperties}.  

\begin{table}
	\caption{Synthetic pulsar properties}
	\label{tab:puslarProperties}
	\begin{center}
	\resizebox{\columnwidth}{!}{%
	\begin{tabular}{lcccc} 
     	Parameter & \multicolumn{4}{c}{Value} \\
		\hline \hline
		 Period(ms) &  \multicolumn{1}{c}{1.102} & \multicolumn{1}{c}{2.218} & \multicolumn{1}{c}{5.218} & \multicolumn{1}{c}{10.870} \\
		   & \multicolumn{1}{c}{18.505} & \multicolumn{1}{c}{61.965} & \multicolumn{1}{c}{126.175} & \multicolumn{1}{c}{286.555} \\
		  & \multicolumn{1}{c}{533.320} & \multicolumn{1}{c}{850.158} & \multicolumn{1}{c}{1657.496} & \multicolumn{1}{c}{2643.410} \\
		  & \multicolumn{1}{c}{3927.013} & \multicolumn{1}{c}{5580.899} & \multicolumn{1}{c}{9964.532}&  \\
		Amplitude & \multicolumn{4}{l}{All pulsars are detectable with a detection} \\ 
		& \multicolumn{4}{l}{significance of $\sim12$ in \texttt{SIGPROC} in the} \\
		& \multicolumn{4}{l}{presence of stationary Gaussian noise when} \\
		& \multicolumn{4}{l}{processed with pipeline H in \texttt{SIGPROC}.} \\
		Duty cycle & \multicolumn{4}{l}{12~$\%$ (fixed)} \\
		Dispersion measure & \multicolumn{4}{l}{68} \\
	\end{tabular}
	}
	\end{center}
\end{table}

A pulse with the profile of pulsar PSR B0833-45 at 1.4 GHz obtained from the EPN-database \citep{b52} with a duty cycle of $12$~$\%$ was injected into all the files.

\subsubsection{Pulsar search pipeline configurations}
\label{sec:Pipelines}

All of the files generated for the experiments described in Table~\ref{tb:Experiments} were processed by both \texttt{SIGPROC} and \texttt{PRESTO}, i.e. twelve different configurations of \texttt{SIGPROC} (see Table~\ref{tb:PipelinesAnalysedSIGPROC}) and eight different configurations of \texttt{PRESTO} (see Table~\ref{tb:PipelinesAnalysedPRESTO}) were used to search every single synthetic file.  Because the aim with this analysis is not to investigate sensitivity as a function of DM, all the files were dedispersed at the correct DM.

\begin{table}
  \caption{The twelve \texttt{SIGPROC} pipeline configurations used to process all the files in this analysis.}
\label{tb:PipelinesAnalysedSIGPROC}
\begin{center}
  \begin{tabular}{  c  c  c  }
     Pipeline & Baseline  & Red-noise mitigation  \\ \hline \hline
     A & Removed & \texttt{default}\\
     B & Removed & \texttt{submn}\\
     C & Removed & \texttt{submjk}\\
     D & Removed & -\\
	 E & Intact & \texttt{default}\\
	 F & Intact & \texttt{submn}\\
	 G & Intact & \texttt{submjk}\\
	 H & Intact & -\\ 
	 I & Moving average filter & \texttt{default}\\
	 J & Moving average filter & \texttt{submn}\\
	 K & Moving average filter & \texttt{submjk}\\
	 L & Moving average filter & -\\
	 
  \end{tabular}
  \end{center}
\end{table}

\texttt{SIGPROC} by default removes the baseline of a dedispersed time series by linearly detrending the time series unless the flag \texttt{-nobaseline} is set when the \texttt{dedisperse} function is called.  In addition to the option available in \texttt{SIGPROC} for detrending the baseline, a 10~s moving average filter was implemented as a second option for normalising the baseline of a dedispersed time series. The red-noise mitigation techniques applied in the different \texttt{SIGPROC} pipelines are described in \S~\ref{sec:SpectrumWhitening}.

To process all the files with \texttt{SIGPROC} the following functions and their associated flags were called:
\begin{enumerate}[label=(\alph*),wide = 0pt, labelwidth = 1.3333em, labelsep = 0.6333em, leftmargin = \dimexpr\labelwidth + \labelsep\relax ]
\item function \texttt{dedisperse} with the flags \texttt{-d}, \texttt{-o} and with/without \texttt{-nobaseline},
\item function \texttt{seek} (number of summed harmonics is 16) with the flags \texttt{-z} and \texttt{-submn}/\texttt{-submn}/\texttt{-submn},
\item function \texttt{best} with flag \texttt{-s8}.
\end{enumerate}
The function \texttt{best} in \texttt{SIGPROC} produce a '.lis' file which was searched for possible candidates based on the SNR of the peaks. 
 
The RFI mask configuration option in Table~\ref{tb:PipelinesAnalysedPRESTO} refers to the RFI mask computed in \texttt{PRESTO} when the \texttt{rfifind} function is called. In this analysis an RFI mask was computed and applied to each synthetic filterbank file at integration intervals of 8 s. An integration time of 8~s was chosen to resemble typical real-time processing intervals. The default values for the time and frequency rejection thresholds in the \texttt{rfifind} function was selected. A moving average filter of 10~s was also implemented as a processing step in the \texttt{PRESTO} pipelines. Lastly, details of the red-noise mitigation technique in \texttt{PRESTO} can be found in \S~\ref{sec:SpectrumWhitening}. 
 
To process all the files with \texttt{PRESTO} the following functions and their associated flags were called:
\begin{enumerate}[label=(\alph*),wide = 0pt, labelwidth = 1.3333em, labelsep = 0.6333em, leftmargin = \dimexpr\labelwidth + \labelsep\relax ]
\item function \texttt{prepdata} with the flags \texttt{-dm}, \texttt{-o} and with/without the flag \texttt{-mask} 
\item function \texttt{realfft},
\item function \texttt{zapbirds} with the flags \texttt{-zap} and \texttt{-zapfile},
\item function \texttt{accelsearch} with the flags \texttt{-sigma~1.0},\\ \texttt{-flo~0.1}, \texttt{-zmax~0} (acceleration searching was turned off by setting the flag \texttt{-zmax 0}) and \texttt{-numharm 16} (i.e. the number of summed harmonics is 16).
\end{enumerate}
The \texttt{accelsearch} function in \texttt{PRESTO} produce an ACCEL file which was searched for possible candidates based on the Gaussian significance of the peaks under the assumption of pure white noise.

\begin{table}
  \caption{The eight \texttt{PRESTO} pipeline configurations used to process all the files in this analysis.}
\label{tb:PipelinesAnalysedPRESTO}
\begin{center}
  \begin{tabular}{  c  c  c  c  }
	 Pipeline & RFI  & Moving average & Red-noise \\ 
     & mask & filter & mitigation  \\ \hline \hline
     A & X & X & X \\
     B & X & X & $\checkmark$ \\
     C & X & $\checkmark$ & X \\
     D & X & $\checkmark$ & $\checkmark$ \\
	 E & $\checkmark$ & X & X \\
	 F & $\checkmark$ & X & $\checkmark$ \\
	 G & $\checkmark$ & $\checkmark$ & X \\
	 H & $\checkmark$ & $\checkmark$ & $\checkmark$ \\
  \end{tabular}
  \end{center}
\end{table}

\section{Results}

\label{sec:Results}

The results are organised according to the aims set forth in the introduction (see \S~\ref{sec:Introduciton}) of this paper.

The heuristics used for quantifying the results are:
\begin{enumerate}[label=(\alph*),wide = 0pt, labelwidth = 1.3333em, labelsep = 0.6333em, leftmargin = \dimexpr\labelwidth + \labelsep\relax ]%
\item a signal is considered a candidate if its detection significance is greater than the default detection threshold in \texttt{SIGPROC} or \texttt{PRESTO},
\item injected pulsars are considered discovered when (a) holds true AND if the difference between the periods of the discovered and injected signals are less than the allowed $error$, i.e.: \\ $|\text{Period}_\textit{discovered}-\text{Period}_\textit{injected}|\le error$, where 
\begin{itemize}[label={}]
\item $error=0.02$ ms if period$\le10$ ms  
\item $error=0.20$ ms if $10$ ms$<$period$\le100$ ms  
\item $error=2.00$ ms if $100$ ms$<$period$\le1000$ ms  
\item $error=20.0$ ms if $1000$ ms$<$period$\le10000$ ms  
\end{itemize}
\item the discoveries from (b) are validated by visual inspection of their folded profiles produced by folding the inverse Fourier transform of their whitened spectra at the detected periods. At this stage a detection is rejected if the folded profile does not resemble a real pulsar,  
\item determining whether a pulsar was detected or not the non-fundamental harmonics were not considered,
\item harmonically related candidates are removed and,
\item only candidates with $1\text{ ms} \le \text{period} \le 10 \text{ s}$ are considered.
\end{enumerate}

For the direct \texttt{SIGPROC} and \texttt{PRESTO} comparisons the exact same files were searched by both routines. 

The sensitivity of a pipeline refers to the ability of the pipeline to detect the fifteen randomly injected pulsars expressed as a percentage. The number of false positives detected per true positive is the total number of false positive candidates detected across all hundred files divided by the number of true positives detected.

Note that all the results presented here should be interpreted as per DM.

\subsection{Non-stationary Gaussian noise and RFI}
\begin{figure*}
      \begin{subfigure}[b]{0.48\linewidth}
          \includegraphics[width=\linewidth]{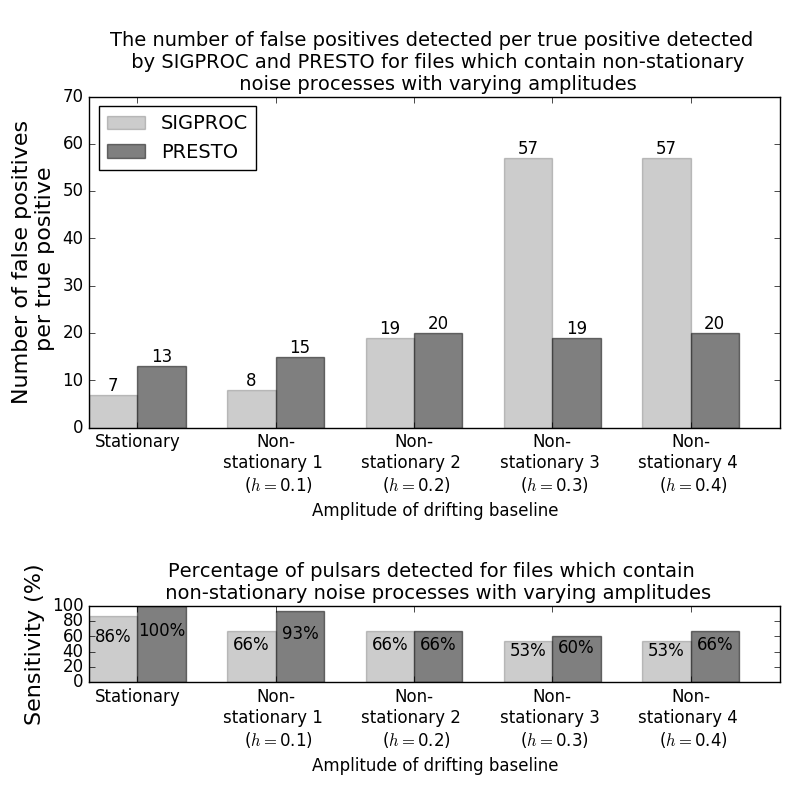}
          \caption{No RFI}
          \label{fig:Question1_V2}
      \end{subfigure}
      ~
      \begin{subfigure}[b]{0.478\linewidth}
          \includegraphics[width=\linewidth]{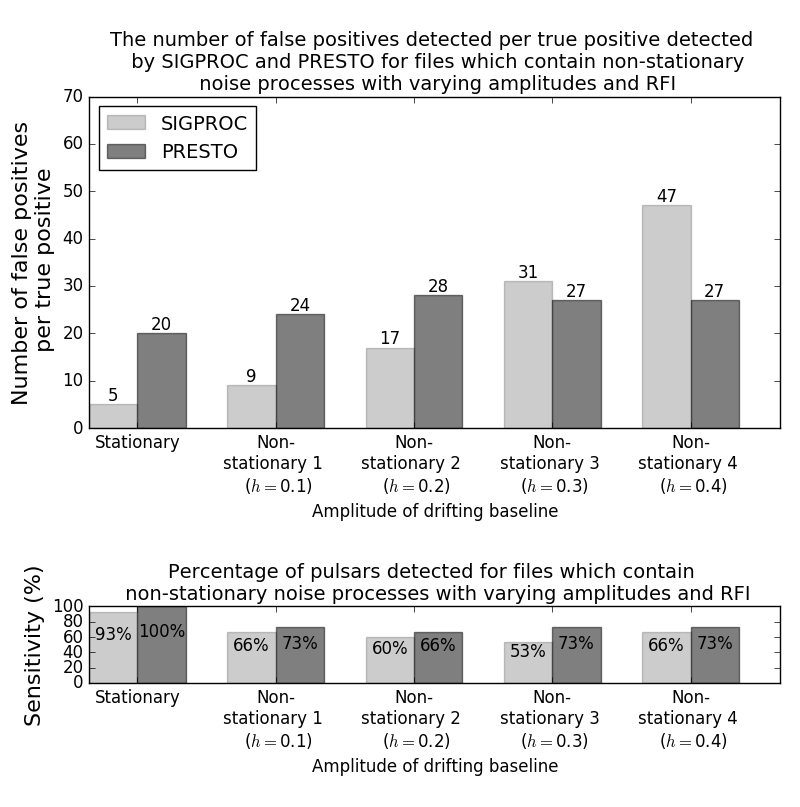}
          \caption{RFI}
          \label{fig:Question1plusRFI_V2}
      \end{subfigure}
  \caption{The performance of \texttt{SIGPROC} and \texttt{PRESTO} for processing files which contain either stationary noise or non-stationary noise with varying amplitudes (i.e. different values for $h$): (a) without RFI (see experiments 1 and 2 in Table~\ref{tb:Experiments}); (b) with RFI (see experiments 3 and 4 in Table~\ref{tb:Experiments}). (\texttt{SIGPROC} pipeline: default baseline subtraction $\rightarrow$ default red-noise removal. \texttt{PRESTO} pipeline: RFI mask $\rightarrow$ baseline not subtracted $\rightarrow$ default red-noise removal.)}
  \label{fig:Question1}
\end{figure*}

The results of processing the synthetic files from the emulated blind surveys (see experiments 1-4 in Table~\ref{tb:Experiments}) with the default pulsar search pipelines of \texttt{SIGPROC} and \texttt{PRESTO} are shown in Figure~\ref{fig:Question1}. Note, the metric used in this section to express the performance of each pipeline is the number of false positives detected for every true positive detected across all 100 files for each experiment.

The number of false positives per true positive detected by \texttt{SIGPROC} increases approximately proportionally with a linear increase in the amplitude of the non-stationary noise (see Figure~\ref{fig:Question1_V2}). This trend is also visible in Figure~\ref{fig:Question1plusRFI_V2} when RFI is injected. Hence, the default \texttt{SIGPROC} pipeline is very sensitive to non-stationary noise.

The number of false positives detected per true positive by \texttt{PRESTO} is unaffected by the type and amplitude of the noise process present in the files, i.e. the number of false positives detected per true positive is almost constant irrespective of the amplitude of the non-stationary noise (see Figure~\ref{fig:Question1_V2}). However, the number of false positives detected per true positive by \texttt{PRESTO} is slightly higher when weak RFI is present (see \S~\ref{sec:RFIDetectionAndMitigation}) compared to when no RFI is injected. 

The sensitivity of \texttt{SIGPROC} and \texttt{PRESTO} can be seen in Figure~\ref{fig:Question1_V2} to decrease by at least $20$~$\%$ and $7$~$\%$ respectively in the presence of non-stationary noise compared to the stationary noise case. The $20$~$\%$ and $7$~$\%$ losses recorded in sensitivity were averaged over all the pulse periods; however, the long period pulsars were much more affected. Note, the amplitudes of the injected pulsars were chosen such that they are detectable at a SNR of $\sim12$ in the presence of white noise when processed with pipeline H in \texttt{SIGPROC} (see Table~\ref{tab:puslarProperties}); however, the addition of any (significant) amount of non-stationary noise rendered the pulsars undetectable. Consequently, there is no correlation visible between sensitivity loss and the non-stationary noise amplitude.

There is no correlation between the loss in sensitivity of \texttt{SIGPROC} and the amplitude of the non-stationary noise when weak RFI is present. Interestingly, the presence of weak RFI leads to an increase in the sensitivity of \texttt{SIGPROC} for the stationary noise case with RFI compared to the stationary noise files without RFI. Similarly, there is an increase in sensitivity for the non-stationary 4 case when RFI is present compared to the no RFI case.

Comparing the sensitivity of \texttt{PRESTO} in Figure~\ref{fig:Question1_V2} to the sensitivity in Figure~\ref{fig:Question1plusRFI_V2} it becomes apparent that \texttt{PRESTO} is not sensitive to weak RFI. The highest sensitivity attainable with \texttt{PRESTO} for files containing weak RFI and non-stationary noise is $73$~$\%$; furthermore, a direct comparison reveals that \texttt{PRESTO}'s sensitivity is on average $11$~$\%$ better than \texttt{SIGPROC}'s sensitivity.

\subsection{Spectrum whitening methods}
\begin{figure}
      \centering
      \includegraphics[width=\linewidth]{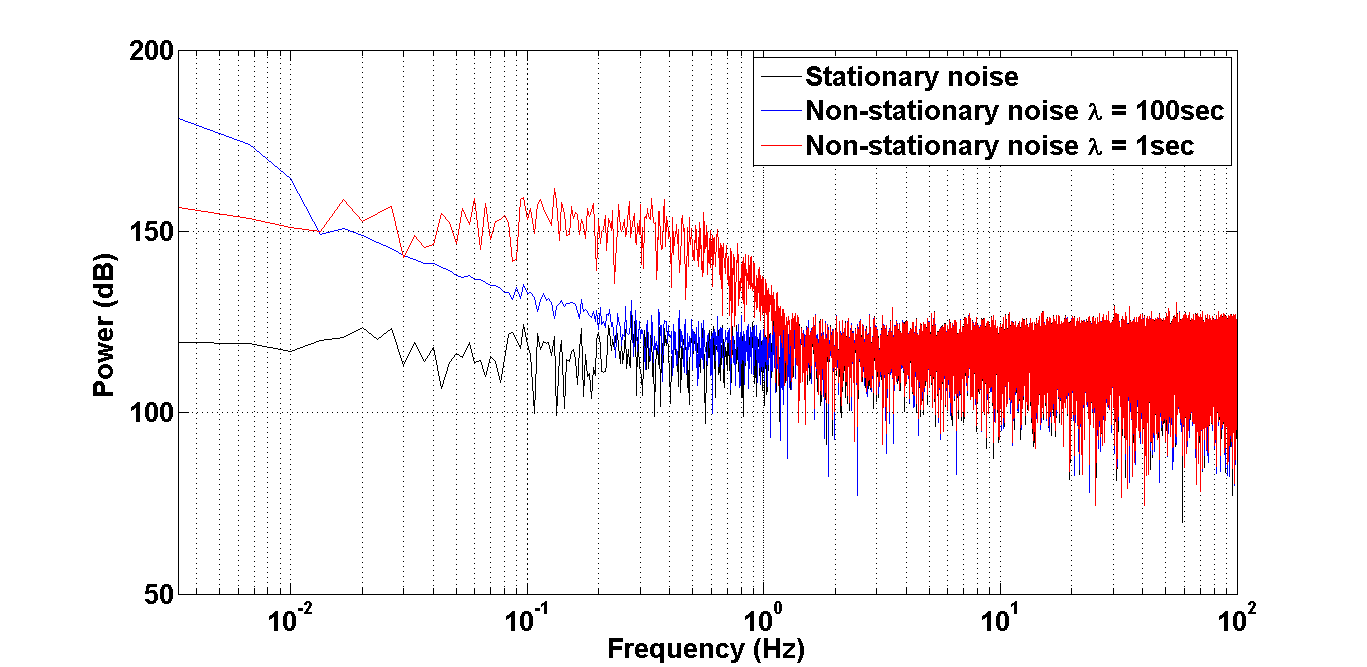}
      \caption{The power spectrum density for different noise processes.}
      \label{fig:ColouredNoise300sec}
\end{figure}

The power versus log-frequency plot in Figure~\ref{fig:ColouredNoise300sec} shows the power spectrum density of two non-stationary Gaussian noise processes with correlation lengths $\lambda=1$~s (red) and $\lambda=100$~s (blue) as well as for a stationary Gaussian noise process (black). It is evident from  Figure~\ref{fig:ColouredNoise300sec} that the power spectrum density of a non-stationary process diverges from the desired flat power spectrum density of a stationary white noise process as the correlation length of the non-stationary process shortens.

\begin{figure*}
  \begin{subfigure}[b]{0.48\linewidth}
      \includegraphics[width=\linewidth]{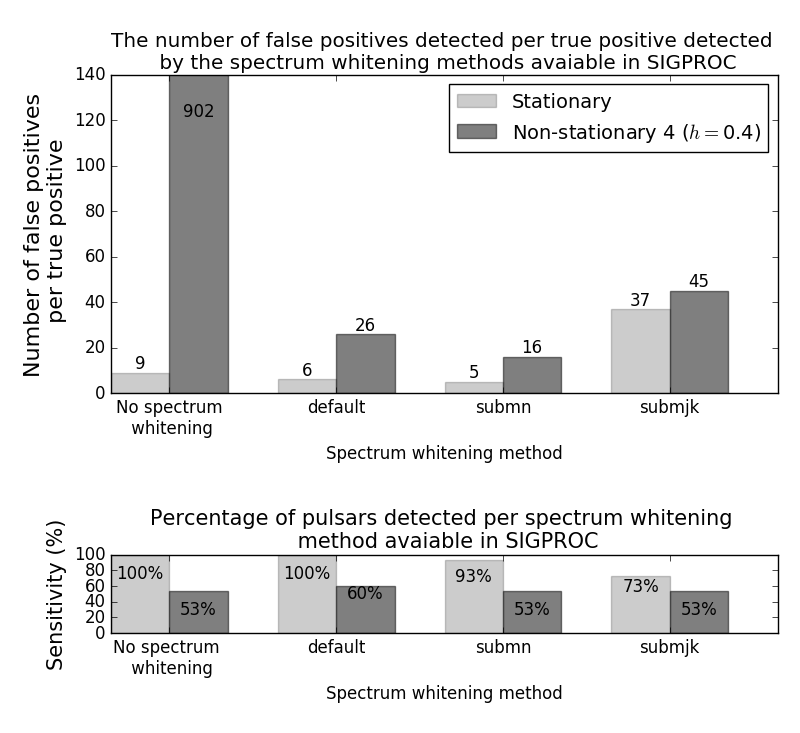}
      \caption{\texttt{SIGPROC}}
      \label{fig:Question2_Sig_V2}
  \end{subfigure}
  ~
  \begin{subfigure}[b]{0.48\linewidth}
      \includegraphics[width=\linewidth]{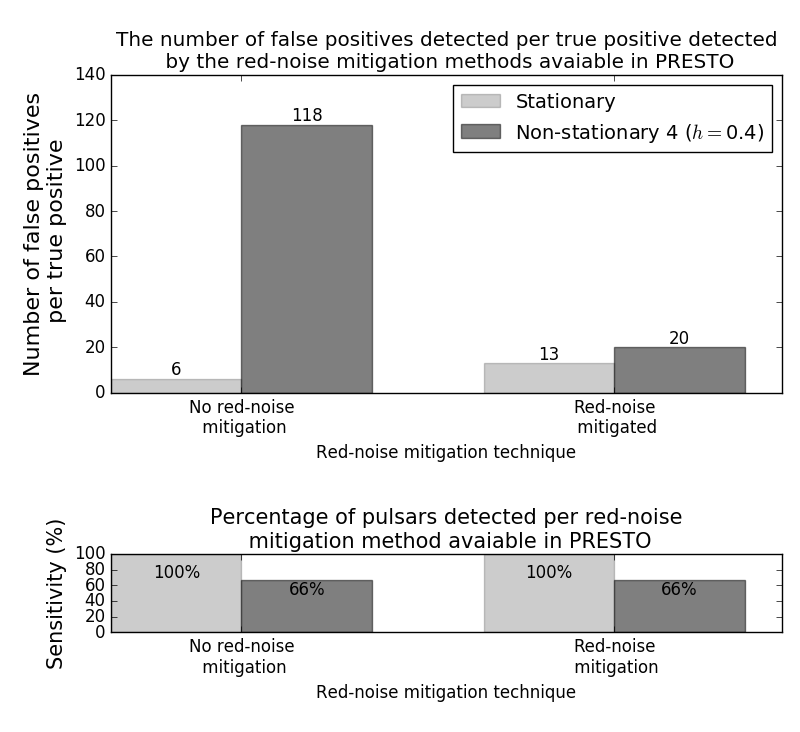}
      \caption{\texttt{PRESTO}}
      \label{fig:Question2_Presto_V2}
  \end{subfigure}
\caption{The performance of the red-noise mitigation methods available in (a) \texttt{SIGPROC} and (b) \texttt{PRESTO} for processing files which contain either stationary noise (see experiment 1 in Table~\ref{tb:Experiments}) or non-stationary noise (see experiment 2 in Table~\ref{tb:Experiments}). No RFI were present in the files analysed. (\texttt{SIGPROC} pipeline: baseline intact $\rightarrow$ red-noise removal methods. \texttt{PRESTO} pipeline: RFI mask $\rightarrow$ baseline not subtracted $\rightarrow$ red-noise removal method.)}
    \label{fig:Question2}
\end{figure*}

With Figure~\ref{fig:ColouredNoise300sec} in mind, four spectrum whitening methods (see \S~\ref{sec:SpectrumWhitening}) were assessed and the results can be seen in Figure~\ref{fig:Question2}. The spectrum whitening techniques in both \texttt{SIGPROC} and \texttt{PRESTO} reduce the number of false positives detected per true positive significantly compared to the case when no spectrum whitening is applied. In the presence of non-stationary noise the sensitivity of \texttt{SIGPROC} improved slightly from $53$~$\%$ to $60$~$\%$ when the default spectrum whitening method was applied but the other methods had no effect on sensitivity (see Figure~\ref{fig:Question2_Sig_V2}). The sensitivity of \texttt{PRESTO} is unchanged when the spectrum whitening method is applied compared to no spectrum whitening.

\subsection{De-trending the data before processing}

\begin{figure*}
	\centering
      \begin{subfigure}[b]{0.48\textwidth}
          \includegraphics[width=\textwidth]{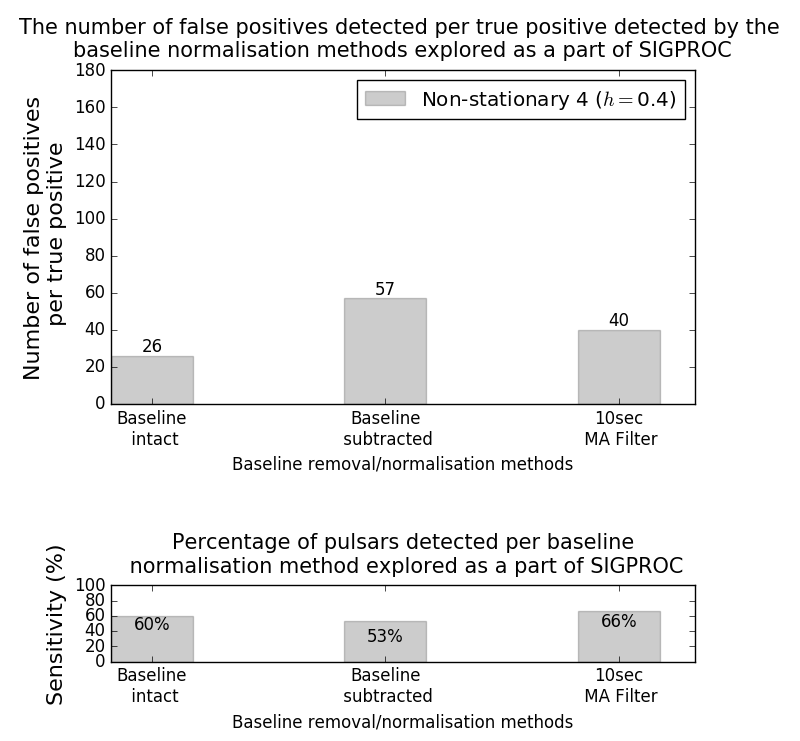}
          \caption{\texttt{SIGPROC}}
          \label{fig:Question3_Sig_V2}
      \end{subfigure}
      ~
      \begin{subfigure}[b]{0.48\textwidth}
          \includegraphics[width=\textwidth]{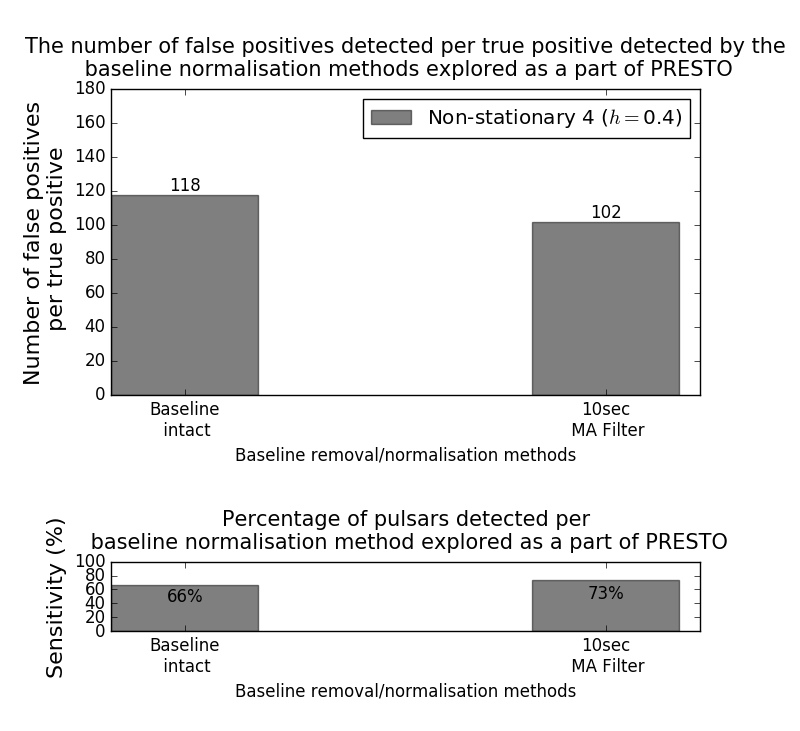}
          \caption{\texttt{PRESTO}}
          \label{fig:Question3_Presto_V2}
      \end{subfigure}
  \caption{The performance of the time-domain baseline normalisation methods available in (a) \texttt{SIGPROC} and (b) \texttt{PRESTO} for processing files which contain non-stationary noise (see experiment 2 in Table~\ref{tb:Experiments}). No RFI was present in the files analysed. (\texttt{SIGPROC} pipeline: three time-domain baseline normalisation methods $\rightarrow$ default red-noise removal. \texttt{PRESTO} pipeline: No RFI mask $\rightarrow$ baseline normalisation method $\rightarrow$ no red-noise removal.)}
  \label{fig:Question3}
\end{figure*}

De-trending the baseline in \texttt{SIGPROC} with either a 10~s moving average filter or the built-in de-trending method led to an increase in the  number of false positives detected per true positive compared to when the baseline was left intact (see Figure~\ref{fig:Question3_Sig_V2}). However, the 10~s moving average filter did improve the sensitivity by $6$~$\%$.

De-trending the baseline in \texttt{PRESTO} with a 10~s moving average filter reduced the number of false positives detected per true positive and increased \texttt{PRESTO's} sensitivity with $7$~$\%$  (see Figure~\ref{fig:Question3_Presto_V2}).

These results hint at the improved sensitivity attainable when the file contains both a baseline with long correlations and a slowly pulsating pulsar, i.e. removing the baseline significantly improves the sensitivity of detecting slow pulsars. This fact is highlighted with the postcard plots in sections \S~\ref{sec:PostcardPresto} and \S~\ref{sec:PostcardSigproc}, described later in the paper. 

\subsection{RFI detection and mitigation methods}
\label{sec:RFIDetectionAndMitigation}

\begin{figure*}
      \begin{subfigure}[b]{0.32\linewidth}
          \includegraphics[width=\linewidth]{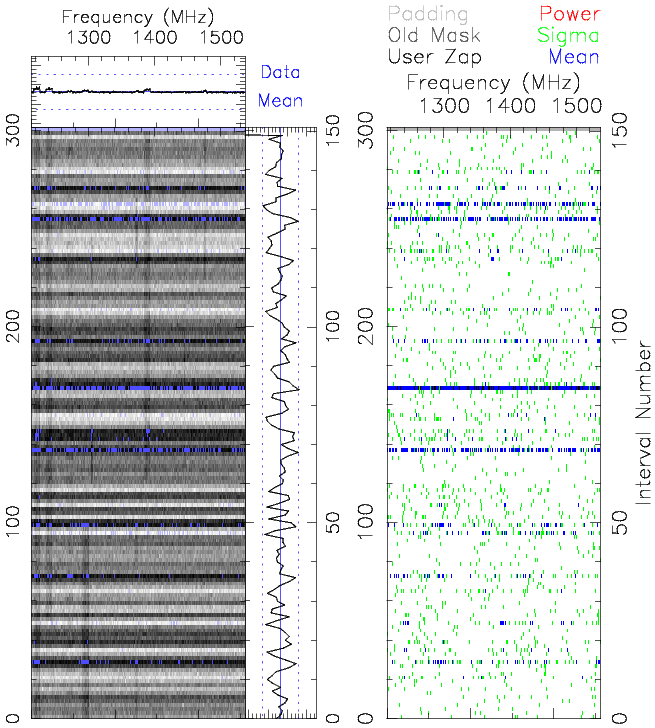}
          \caption{2~s integration}
          \label{fig:RFI2sec}
      \end{subfigure}
      ~
      \begin{subfigure}[b]{0.32\linewidth}
          \includegraphics[width=\linewidth]{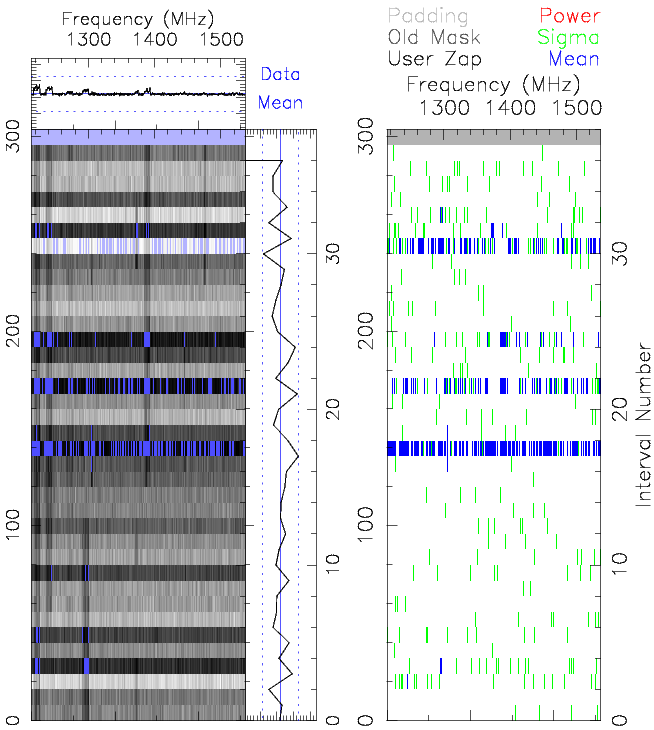}
          \caption{8~s integration}
          \label{fig:RFI8sec}
      \end{subfigure}
      ~
      \begin{subfigure}[b]{0.32\linewidth}
          \includegraphics[width=\linewidth]{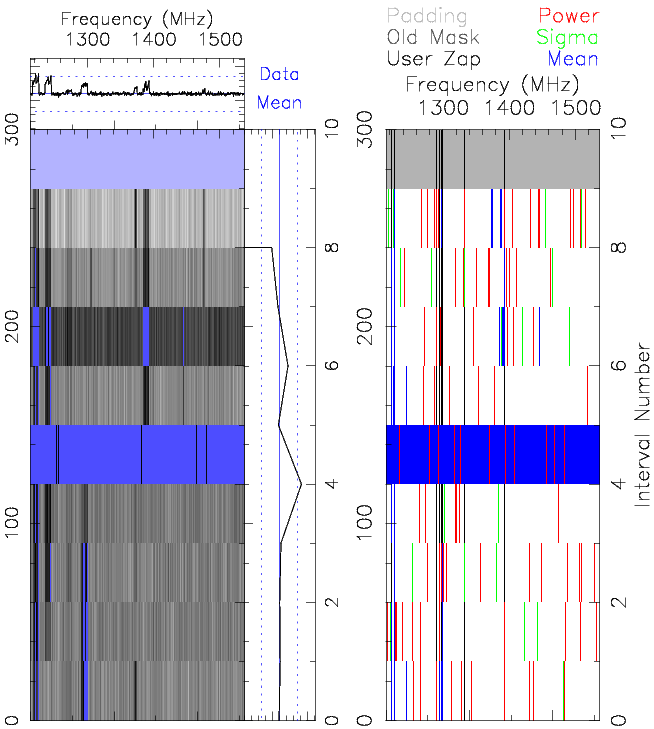}
          \caption{30~s integration}
          \label{fig:RFI30sec}
 \end{subfigure}
  \caption{RFI masks created with \texttt{PRESTO}'s \texttt{rfifind} function. The plots are for the same file but created with different integration times. The \texttt{-timesig} threshold was set to three and the \texttt{-freqsig} threshold was set to eight for the \texttt{rfifind} function.}
 \label{fig:RFIintegrationTimes}
\end{figure*}

RFI masks created with \texttt{PRESTO}'s \texttt{rfifind} function, for the same file, are depicted in Figure~\ref{fig:RFIintegrationTimes}. The masks differ with respect to the integration times used to create them.  With the plots in Figure~\ref{fig:RFIintegrationTimes} we show that the RFI we injected, although visible, is weak compared to the amplitude of the non-stationary baseline.

The default integration length of 30~s is most successful at detecting the actual injected RFI (see Figure~\ref{fig:RFI30sec}). The two masks created with shorter integration times mostly flagged the maxima of the non-stationary baseline as opposed to the actual injected RFI (see Figure~\ref{fig:RFI2sec} and Figure~\ref{fig:RFI8sec}).

RFI was injected such that $12$~$\%$ of all the samples in the data are affected. The 2~s mask in Figure~\ref{fig:RFI2sec} found $6.868$~$\%$ of the 2~s intervals to be affected by RFI, the 8~s mask in Figure~\ref{fig:RFI8sec} found $6.887$~$\%$ of the 8~s intervals to be affected by RFI and the 30~s mask found that $16.895$~$\%$ of the 30~s intervals are affected by RFI. 

\begin{figure}
\centering
\includegraphics[width=0.48\textwidth]{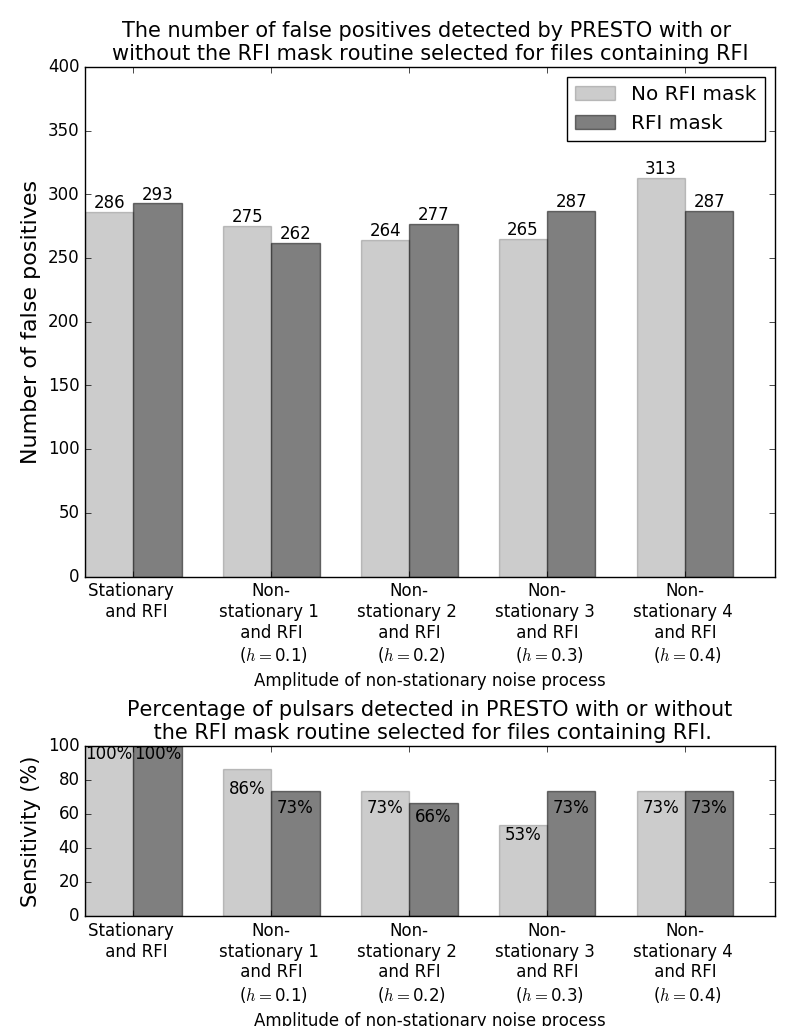}
\caption{The efficacy of the RFI detection and masking routine in \texttt{PRESTO} when processing files which contain either non-stationary noise (see experiment 2 in Table~\ref{tb:Experiments}) or non-stationary noise with weak RFI (see experiment 4 in Table~\ref{tb:Experiments}).  (\texttt{PRESTO} pipeline: RFI Mask Yes/No $\rightarrow$ baseline intact $\rightarrow$ default red-noise removal.)}
\label{fig:Question5}
\end{figure}

The focus of this section is on the real-time detection and mitigation of RFI, hence the decision to investigate the effectiveness of an RFI mask integrated over a few seconds when applied to the synthetic filterbank files. The results of which can be seen in Figure~\ref{fig:Question5}.

The RFI detection and masking routine available in \texttt{PRESTO} had on average little to no effect on both the sensitivity and the number of false positives detected per true positive except for the non-stationary 3 with RFI case where the sensitivity was increased from $53$~$\%$ to $73$~$\%$ when the mask was applied. Greater insight on the effect of applying the RFI detection and mitigation routine in \texttt{PRESTO} to files containing weak RFI can be gained by comparing the results of pipelines A-D to those of pipelines E-H in Figure~\ref{tbl:SensitivityPresto}.
    
\subsection{Variation of detection with signal significance}
\label{sec:SignalSignificance}

The detection significance which pulsars embedded in different noise processes were detected at by the default search pipelines of \texttt{SIGPROC} and \texttt{PRESTO} are depicted in Figure~\ref{fig:Question6} with the colours being representative of the following:
\begin{enumerate}[label=(\alph*),wide = 0pt, labelwidth = 1.3333em, labelsep = 0.6333em, leftmargin = \dimexpr\labelwidth + \labelsep\relax ]%
\item Green square: an injected pulsar was detected (the detection significance is printed in the square), 
\item Orange square: an injected pulsar is amongst the detected signals but is not considered a candidate because its detection significance is not above the default threshold value, 
\item Red square: signifies that an injected pulsar was missed,
\item Grey square: signifies that an injected pulsar was detected but is not considered a candidate because it has an abnormally high detection significance.
\end{enumerate}
Each box in Figure~\ref{fig:Question6} show a single instantiation of a pulsar/non-stationary baseline pair that was searched by both \texttt{SIGPROC} and \texttt{PRESTO}.

The left hand side of both matrices in Figure~\ref{fig:Question6} are populated with detections, whereas the right hand sides are predominantly populated with misses. Hence, long-period pulsars embedded in non-stationary noise processes across a range of correlation lengths are missed by both the default \texttt{SIGPROC} and \texttt{PRESTO} pulsar search pipelines.

The detection significance at which pulsars with periods greater than 50 ms are detected decreases as the correlation length of the non-stationary noise shortens, whereas the detection significance of fast-period pulsars are unaffected by the correlation length of the non-stationary noise.

The results in Figure~\ref{fig:Question6} portray single-trials for near-threshold signals which are very sensitive to the noise realisation used. To help understand the average and variance associated with the detection significance of these signals, we injected a pulsar with a period of 0.126~s in an ensemble of 20 noise realisations each with the same correlation length of 1~s and amplitude $h=0.4$ (see Equation~\ref{eq:Lowpass}). This additional experiment showed that the average SNR at which the pulsar was detected in \texttt{SIGPROC} is 9 with a standard deviation of 1.45 compared to the SNR of 12.1 at which the pulsar is detected when embedded in stationary Gaussian noise. Similarly, the average Gaussian significance of the detected pulsar in \texttt{PRESTO} is 6.621 with a standard deviation of 1.14 compared to the stationary Gaussian noise case of 7.7. Consequently, the results for this particular combination of period and correlation length that are plotted in Figure~\ref{fig:Question6}, Figure~\ref{tbl:SensitivityPresto}, Figure~\ref{tbl:SensitivitySIGPROC1} and Figure~\ref{tbl:SensitivitySIGPROC2} would show very little variability had there been more realisations of the same experiments. Multiple repetitions of this experiment over all combinations is extremely time costly and has not been attempted here. We have however computed similar standard deviations for other combinations of period and length scale (e.g periods of 5~s, 0.01~s and 0.002~s, and $\lambda$s of 1~s, 0.01~s and 100~s), using small numbers of realizations (5 to 20). We find that the standard deviation in the experiments with no injected RFI remains similar to the measurements above, whereas the cases with RFI show increased variance, with measured standard deviations of between 2 and 3. Although this will have an effect on a case by case basis, the overall statistical picture can be interpreted. 

\begin{figure*}
  \centering
      \begin{subfigure}[b]{0.9\textwidth}
          \includegraphics[width=\textwidth]{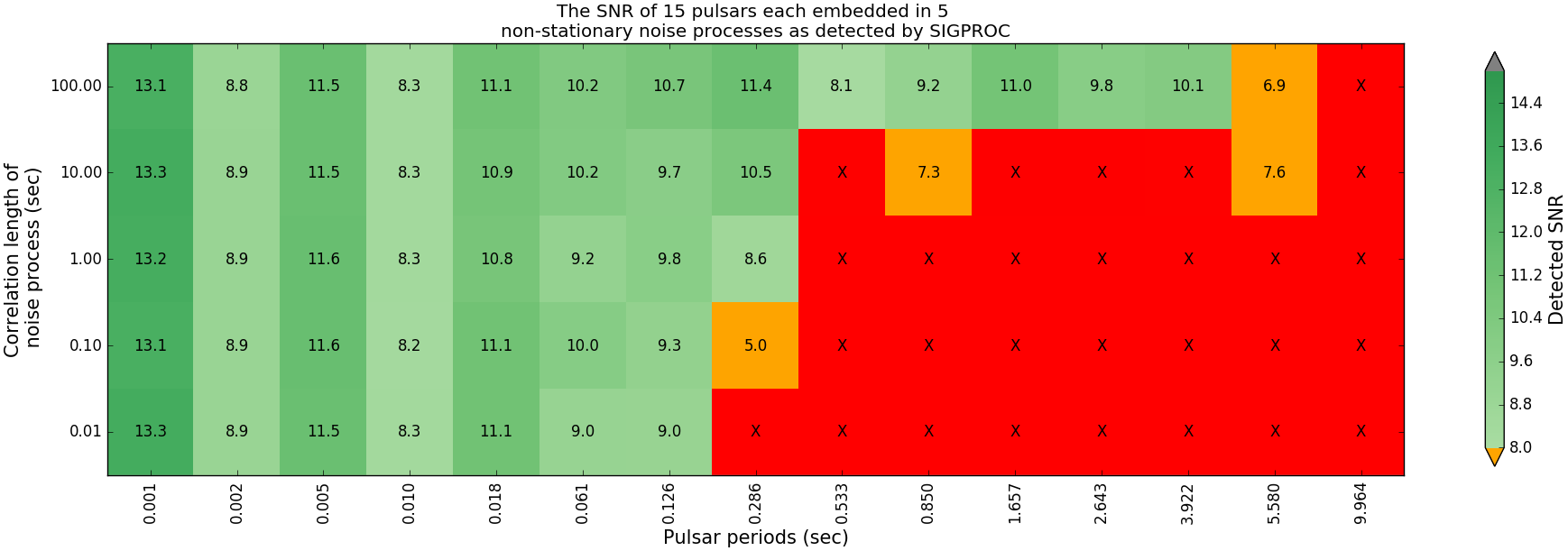}
          \caption{\texttt{SIGPROC}}
          \label{fig:Question6_Sig}
      \end{subfigure}
      
      \begin{subfigure}[b]{0.9\textwidth}
          \includegraphics[width=\textwidth]{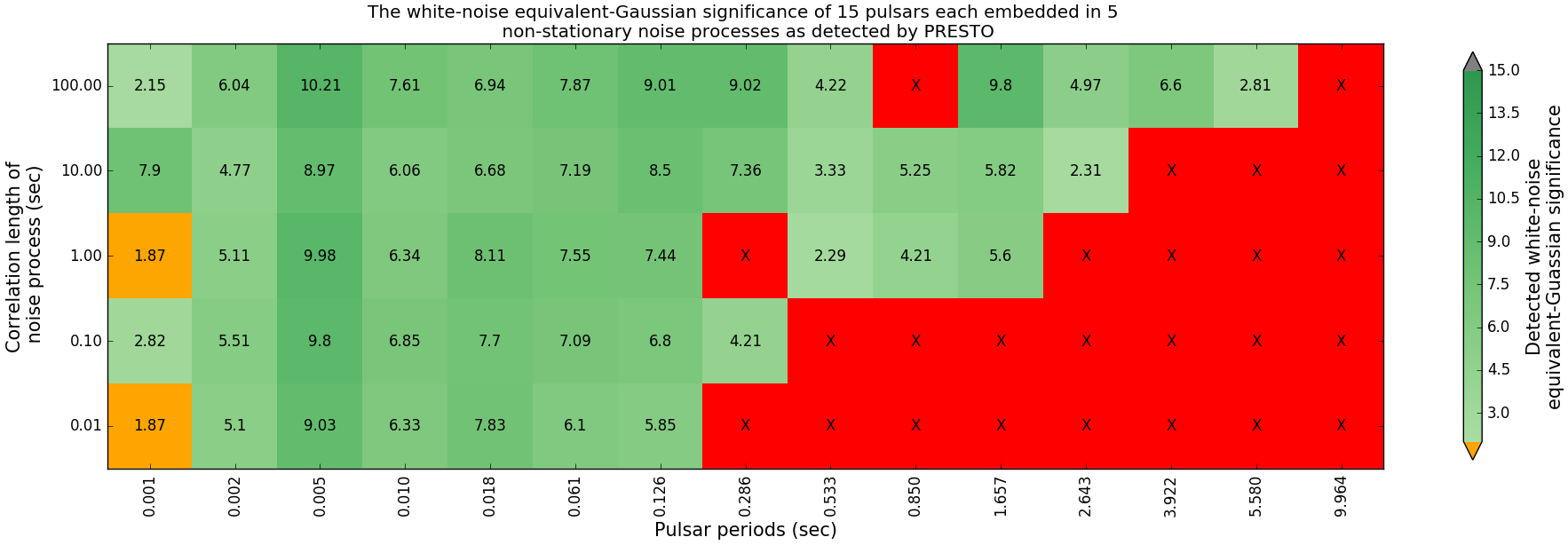}
          \caption{\texttt{PRESTO}}
          \label{fig:Question6_Presto}
      \end{subfigure}
  \caption{The detection significance (values in black) at which 15 pulsars with different periods were detected (green squares) in files containing non-stationary noise processes with a relative amplitude of $h=0.4$ and five different correlation lengths (see experiment 5 in Table~\ref{tb:Experiments}). 
  Each box represents a single instantiation of a pulsar/non-stationary baseline pair that was searched by both \texttt{SIGPROC} and \texttt{PRESTO}. The red squares represent missed pulsars and the orange squares represent detected pulsars with detection significances below the default threshold levels. (a) The results for files processed with the default pipeline in \texttt{SIGPROC}; (b) The results for files processed with the default pipeline in \texttt{PRESTO}. }
  \label{fig:Question6}
\end{figure*}

It is evident from Figure~\ref{fig:Question6_Sig} and Figure~\ref{fig:Question6_Presto} that the default search pipeline in \texttt{PRESTO} is better at finding pulsars of various periods embedded in different non-stationary noise processes compared to \texttt{SIGPROC}. 

\subsection{Sensitivity postcard plots of all the pipelines used to process files with PRESTO}
\label{sec:PostcardPresto}

The sensitivity plots for all the search pipelines explored in \texttt{PRESTO} (see Table~\ref{tb:PipelinesAnalysedPRESTO}) are depicted in Figure~\ref{tbl:SensitivityPresto}.

None of the pipelines in \texttt{PRESTO} are able to detect all the pulsars embedded in the different noise processes. Moreover, most of the pipelines miss the long-period pulsars. The addition of weak RFI, in general, does not alter \texttt{PRESTO}'s ability to find pulsars.

Pipelines A, C, E and G in Figure~\ref{tbl:SensitivityPresto} contain detections with Gaussian significances well in excess of the expected maximum Gaussian significance. We do not consider these outliers as true detections. However, do note that these pipelines all have one thing in common and that is they do not whiten the spectrum. 

The only difference between pipelines A to D and E to H is the application of the RFI masking routine in \texttt{PRESTO}. From the results it appears that the RFI routine attenuates the Gaussian significance of short period pulsars below the detection threshold both in the presence and absence of RFI. 

From these plots it is evident that running a moving average filter to normalise the time-domain data results in improved sensitivity, for example compare pipeline E with G and pipeline F with H. Note, when the moving average filter is applied in conjunction with the red-noise suppression method (see pipeline H in Figure~\ref{tbl:SensitivityPresto}) then more long-period pulsars embedded in non-stationary noise processes with long correlation lengths are detected compared to when only the moving average filter is applied (see pipeline F in Figure~\ref{tbl:SensitivityPresto}).

The pulsar search pipeline D in \texttt{PRESTO} (No RFI mask $\rightarrow$ MA filter $\rightarrow$ red noise mitigation) yields the best results amongst all the set-ups both in the presence and absence of RFI. 

\begin{table*}
  \begin{tabular}{ | m{0.2cm} | c | c | }
    \hline
     & No RFI injected & RFI injected  \\ 
     & (experiment 5 with $h=0.4$) & (experiment 6 with $h=0.4$) \\ \hline   
  	A
    &
	    \begin{minipage}{.4\textwidth}
		  \centering
		  
	      \includegraphics[height=2.00cm]{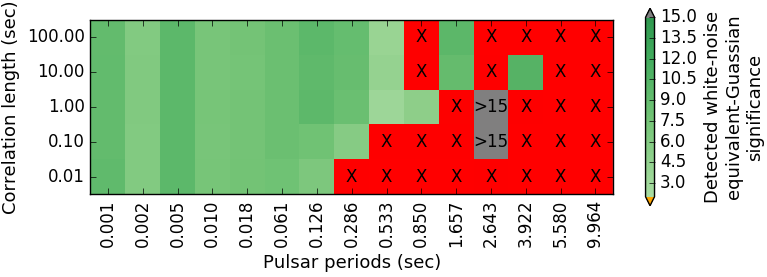} 
	    \end{minipage} 
    & 
	    \begin{minipage}{.4\textwidth}
	      \centering
	      \includegraphics[height=2.00cm]{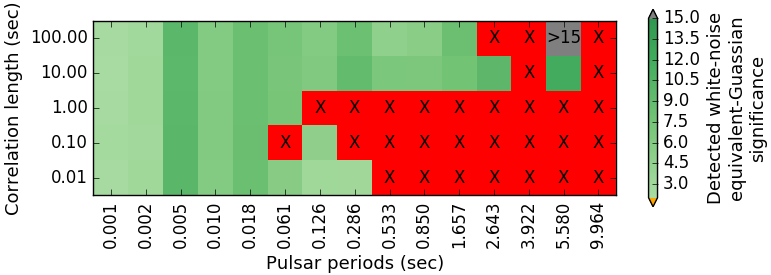}
	    \end{minipage}
      \\   \hline 
   	B
    &
	    \begin{minipage}{.4\textwidth}
   	      \centering
	      \includegraphics[height=2.00cm]{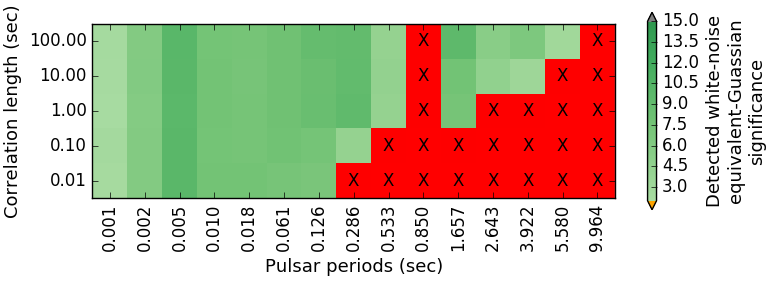}
	    \end{minipage}
    & 
	    \begin{minipage}{.4\textwidth}
	      \centering	    
	      \includegraphics[height=2.00cm]{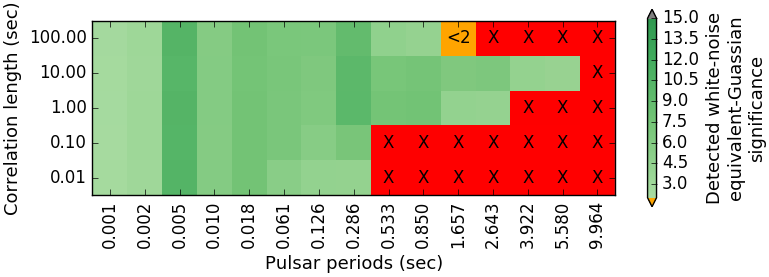}
	    \end{minipage}
    \\ \hline
  	C
    &
	    \begin{minipage}{.4\textwidth}
	      \centering	    
	      \includegraphics[height=2.00cm]{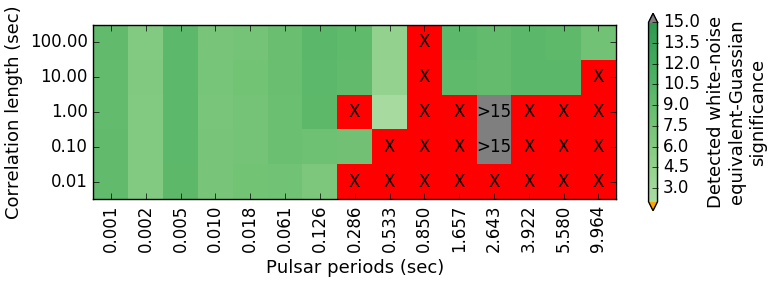} 
	    \end{minipage} 
    & 
	    \begin{minipage}{.4\textwidth}
	      \centering	    
	      \includegraphics[height=2.00cm]{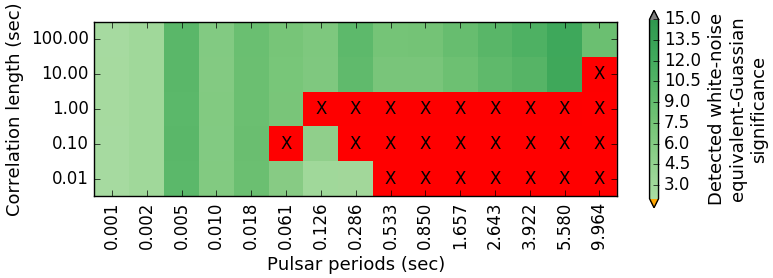}
	    \end{minipage}
      \\   \hline 
   	D
    &
	    \begin{minipage}{.4\textwidth}
	      \centering	    
	      \includegraphics[height=2.00cm]{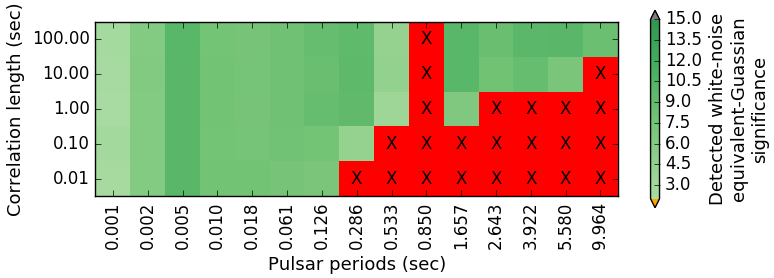}
	    \end{minipage}
    & 
	    \begin{minipage}{.4\textwidth}
	      \centering	     
	      \includegraphics[height=2.00cm]{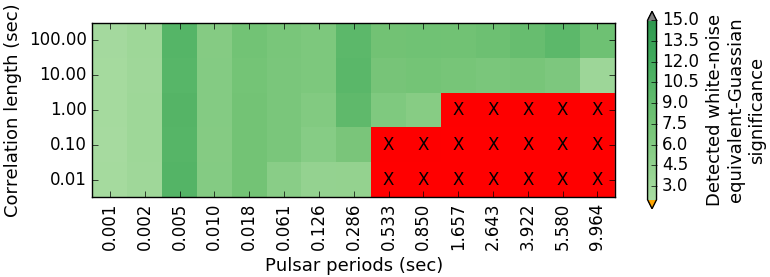}
	    \end{minipage}
    \\ \hline
  	E
    &
	    \begin{minipage}{.4\textwidth}
	      \centering	    
	      \includegraphics[height=2.00cm]{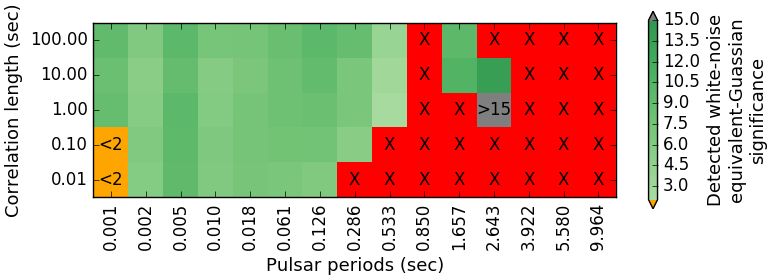} 
	    \end{minipage} 
    & 
	    \begin{minipage}{.4\textwidth}
	      \centering	    
	      \includegraphics[height=2.00cm]{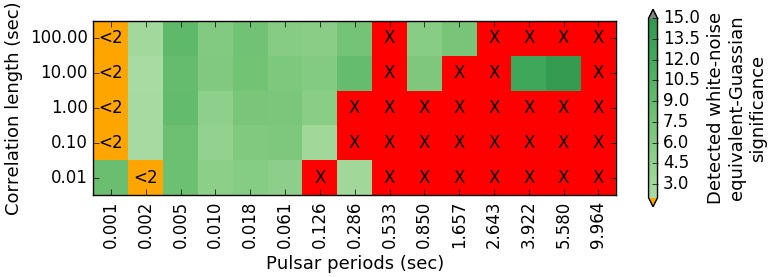}
	    \end{minipage}
      \\   \hline 
   	F
    &      
	    \begin{minipage}{.4\textwidth}
	      \centering	    
	      \includegraphics[height=2.00cm]{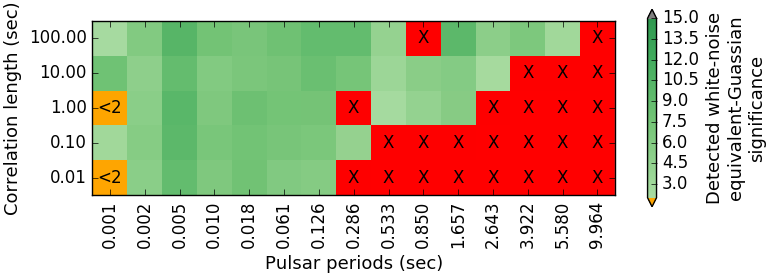}
	    \end{minipage}
    & 
	    \begin{minipage}{.4\textwidth}
	      \centering	    
	      \includegraphics[height=2.00cm]{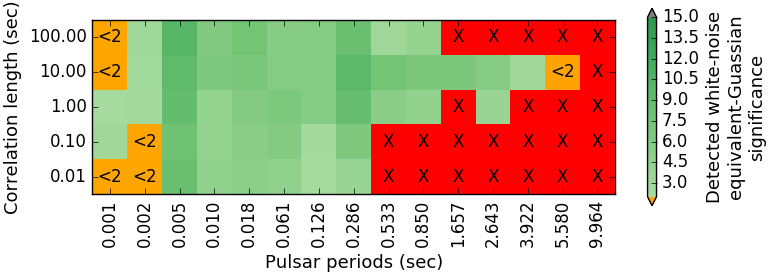}
	    \end{minipage}
    \\ \hline
  	G
    & 
	    \begin{minipage}{.4\textwidth}
	      \centering	    
	      \includegraphics[height=2.00cm]{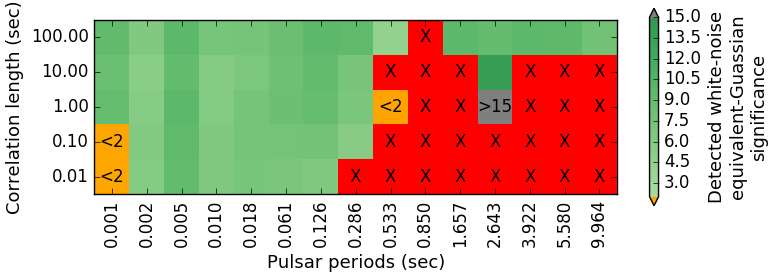} 
	    \end{minipage} 
    & 
	    \begin{minipage}{.4\textwidth}
	      \centering	    
	      \includegraphics[height=2.00cm]{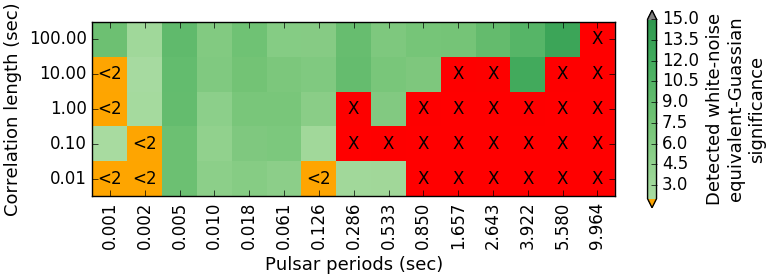}
	    \end{minipage}
      \\   \hline 
   	H
    &
	    \begin{minipage}{.4\textwidth}
	      \centering	    
	      \includegraphics[height=2.00cm]{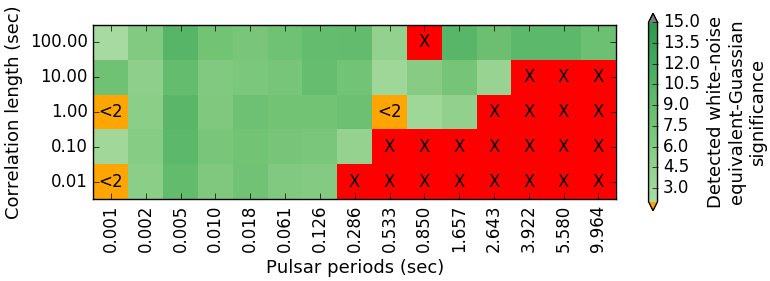}
	    \end{minipage}
    & 
	    \begin{minipage}{.4\textwidth}
	    	      \centering
	      \includegraphics[height=2.00cm]{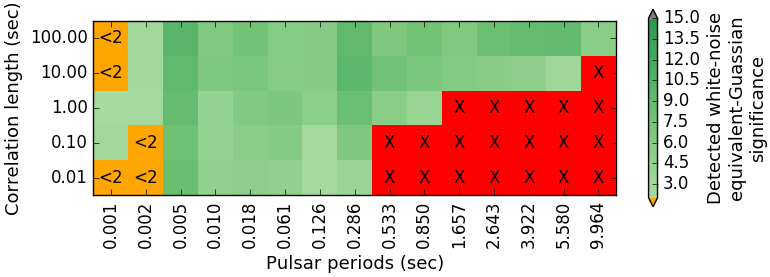}
	    \end{minipage}
    \\ \hline
    
  \end{tabular}
  \captionof{figure}{The Gaussian significance at which pulsars were detected (green squares) after files containing them (see experiments 5 and 6 in Table~\ref{tb:Experiments}) were processed by eight different pipelines in \texttt{PRESTO}. The red squares represent missed pulsars, the orange squares represent detected pulsars with Gaussian significances below the default threshold level of 2 and the grey squares represent detected pulsars with Guassian significances above the average maximum Gaussian significance of 15.}\label{tbl:SensitivityPresto}
\end{table*}

\subsection{Sensitivity postcard plots of all the pipelines used to process files with SIGPROC}
\label{sec:PostcardSigproc}

The sensitivity plots for all the search pipelines explored in \texttt{SIGPROC} (see Table~\ref{tb:PipelinesAnalysedSIGPROC}) are depicted in Figure~\ref{tbl:SensitivitySIGPROC1} and Figure~\ref{tbl:SensitivitySIGPROC2}.

Note, the pulsar with period 0.002218~s is detected below the detection threshold (see orange squares in Figures~\ref{tbl:SensitivitySIGPROC1} and \ref{tbl:SensitivitySIGPROC2}) by almost all of the pipeline configurations in \texttt{SIGPROC} when RFI is present despite the other millisecond pulsars being detected. This pulsar is missed due to the increased variance of the SNR associated with the presence of RFI as explored and explained in section \S~\ref{sec:SignalSignificance}.

It is apparent from pipelines D and H in Figure~\ref{tbl:SensitivitySIGPROC1} and pipeline L in Figure~\ref{tbl:SensitivitySIGPROC2} that not normalising the spectrum results in a lot of pulsars being missed. Furthermore, mostly the long-period pulsars are regularly missed irrespective of the pipeline used in \texttt{SIGPROC}.  

Overall \texttt{PRESTO's} performance across all the pipelines is more consistent when compared the pipelines in \texttt{SIGPROC}.

\begin{table*}
  \begin{tabular}{ | m{0.2cm} | c | c | }
    \hline
     & No RFI injected & RFI injected  \\ 
     & Experiment 5 with $h=0.4$ & Experiment 6 with $h=0.4$ \\ \hline   

  	A
    &
	    \begin{minipage}{.4\textwidth}
   	      \centering
	      \includegraphics[height=2.00cm]{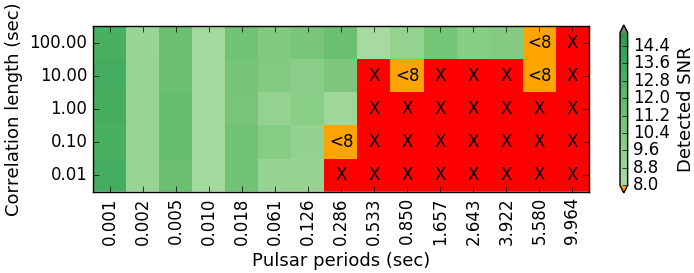} 
	    \end{minipage} 
    & 
	    \begin{minipage}{.4\textwidth}
	      \centering	    
	      \includegraphics[height=2.00cm]{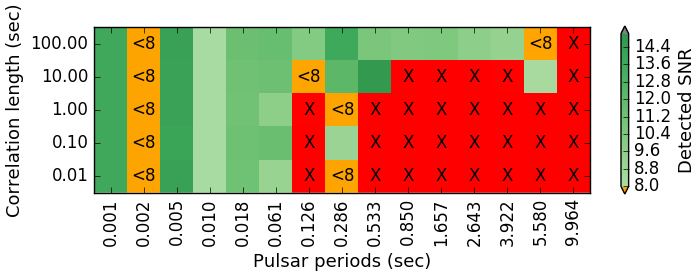}
	    \end{minipage}
      \\   \hline 
   	B
    &
	    \begin{minipage}{.4\textwidth}
	      \centering	    
	      \includegraphics[height=2.00cm]{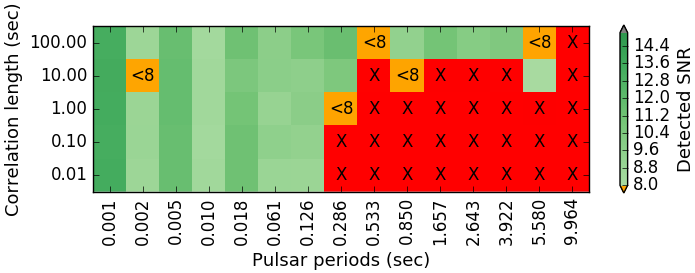}
	    \end{minipage}
    & 
	    \begin{minipage}{.4\textwidth}
	      \centering	    
	      \includegraphics[height=2.00cm]{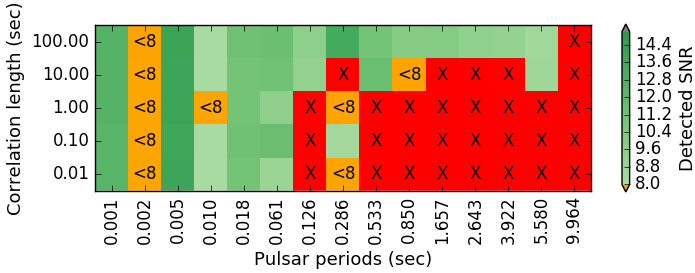}
	    \end{minipage}
    \\ \hline
    
  	C
    &
	    \begin{minipage}{.4\textwidth}
	      \centering	    
	      \includegraphics[height=2.00cm]{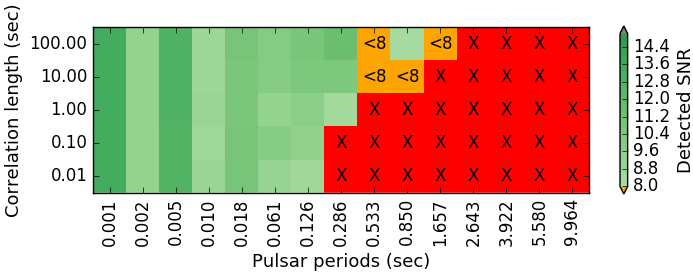} 
	    \end{minipage} 
    & 
	    \begin{minipage}{.4\textwidth}
	      \centering	    
	      \includegraphics[height=2.00cm]{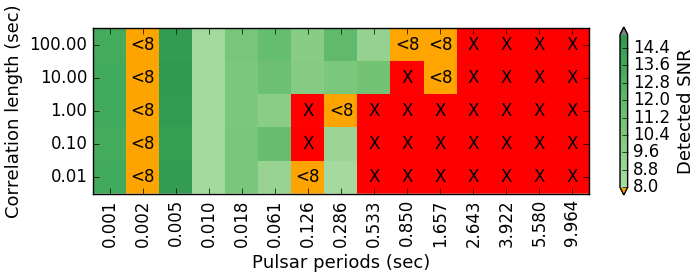}
	    \end{minipage}
      \\   \hline 
   	D
    &
	    \begin{minipage}{.4\textwidth}
	      \centering	    
	      \includegraphics[height=2.00cm]{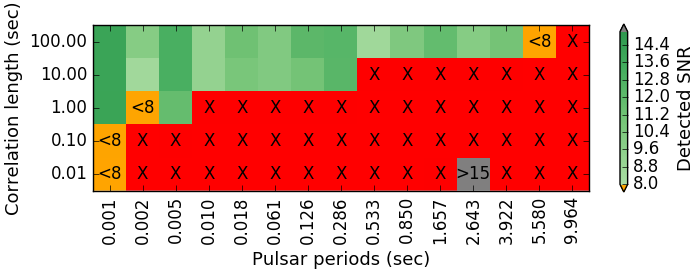}
	    \end{minipage}
    & 
	    \begin{minipage}{.4\textwidth}
	      \centering	    
	      \includegraphics[height=2.00cm]{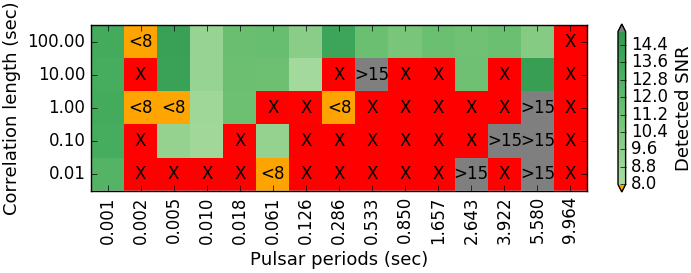}
	    \end{minipage}
    \\ \hline
  	E
    &
	    \begin{minipage}{.4\textwidth}
   	      \centering
	      \includegraphics[height=2.00cm]{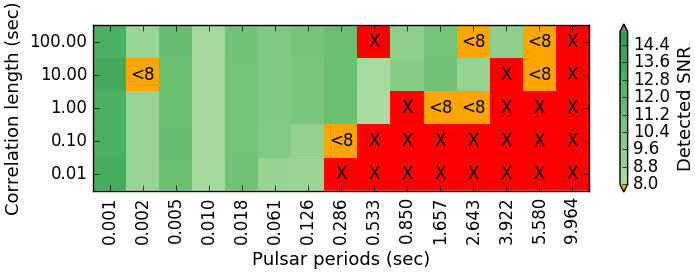} 
	    \end{minipage} 
    & 
	    \begin{minipage}{.4\textwidth}
	      \centering
	      \includegraphics[height=2.00cm]{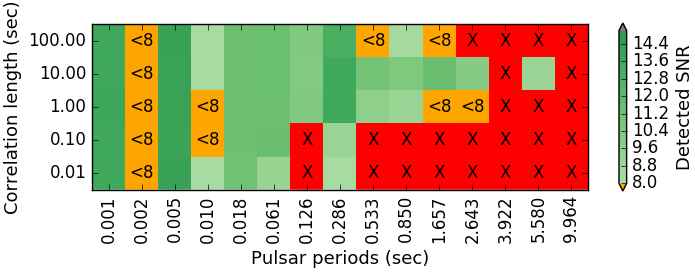}
	    \end{minipage}
      \\   \hline 
    F  
    &
	    \begin{minipage}{.4\textwidth}
	      \centering	    
	      \includegraphics[height=2.00cm]{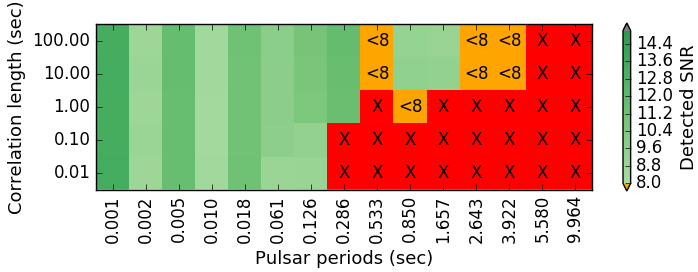}
	    \end{minipage}
    & 
	    \begin{minipage}{.4\textwidth}
	      \centering	    
	      \includegraphics[height=2.00cm]{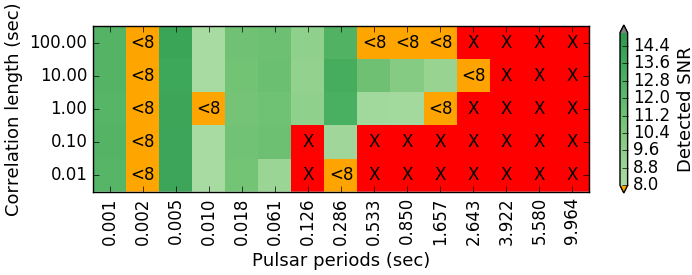}
	    \end{minipage}
    \\ \hline
    
  	G
    &
	    \begin{minipage}{.4\textwidth}
	      \centering	    
	      \includegraphics[height=2.00cm]{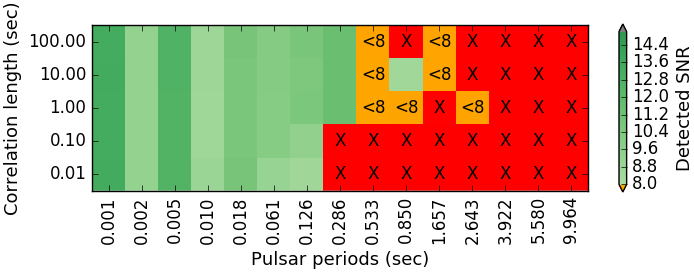} 
	    \end{minipage} 
    &
	    \begin{minipage}{.4\textwidth}
	      \centering	    
	      \includegraphics[height=2.00cm]{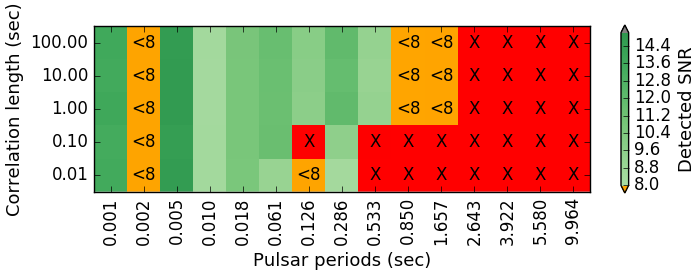}
	    \end{minipage}
      \\   \hline 
   	H
    &
	    \begin{minipage}{.4\textwidth}
	      \centering	    
	      \includegraphics[height=2.00cm]{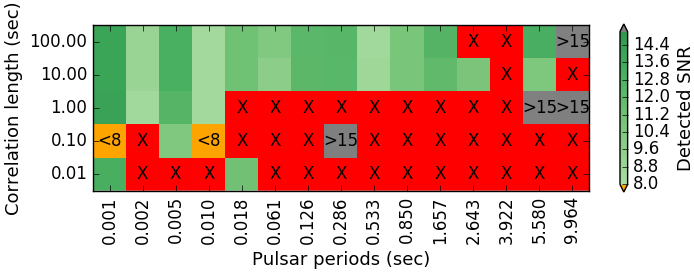}
	    \end{minipage}
    & 
	    \begin{minipage}{.4\textwidth}
	      \centering	    
	      \includegraphics[height=2.00cm]{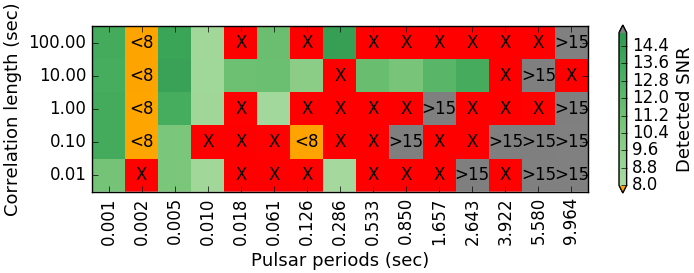}
	    \end{minipage}
    \\ \hline
    
  \end{tabular}
  \captionof{figure}{The SNR at which pulsars were detected after files containing them (see experiments 5 and 6 in Table~\ref{tb:Experiments}) were processed by eight different pipelines in \texttt{SIGPROC}.  The red squares represent missed pulsars, the orange squares represent detected pulsars with SNRs below the default threshold level of 8 and the grey squares represent detected pulsars with SNRs above the average maximum SNR of 15.}\label{tbl:SensitivitySIGPROC1}
\end{table*}

\begin{table*}
  \centering
  \begin{tabular}{ | m{0.2cm} | c | c | }
    \hline
     & No RFI injected & RFI injected  \\ \hline
  	I
    &
	    \begin{minipage}{.4\textwidth}
	      \centering	    
	      \includegraphics[height=2.00cm]{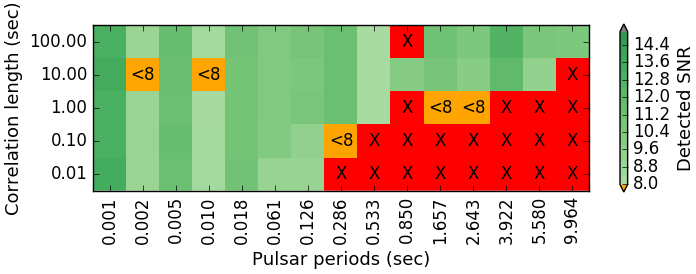} 
	    \end{minipage} 
    & 
	    \begin{minipage}{.4\textwidth}
	      \centering	    
	      \includegraphics[height=2.00cm]{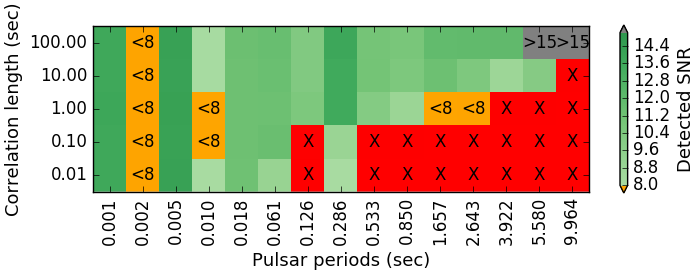}
	    \end{minipage}
      \\   \hline 
   	J
    &
	    \begin{minipage}{.4\textwidth}
	      \centering	    
	      \includegraphics[height=2.00cm]{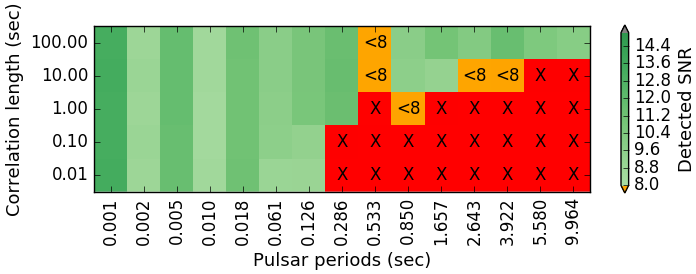}
	    \end{minipage}
    & 
	    \begin{minipage}{.4\textwidth}
   	      \centering
	      \includegraphics[height=2.00cm]{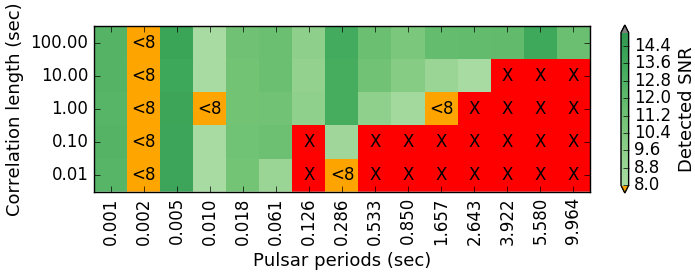}
	    \end{minipage}
    \\ \hline
  	K
    &
	    \begin{minipage}{.4\textwidth}
	      \centering	    
	      \includegraphics[height=2.00cm]{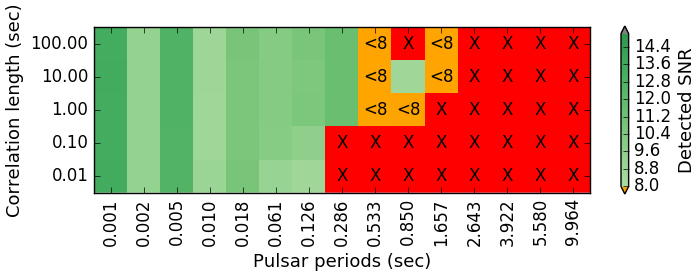} 
	    \end{minipage} 
    & 
	    \begin{minipage}{.4\textwidth}
	      \centering	    
	      \includegraphics[height=2.00cm]{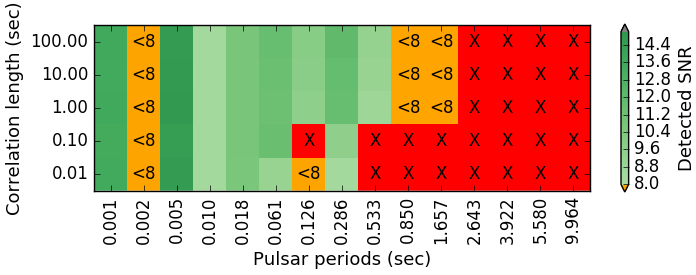}
	    \end{minipage}
      \\   \hline 
   	L
    &
	    \begin{minipage}{.4\textwidth}
	      \centering	    
	      \includegraphics[height=2.00cm]{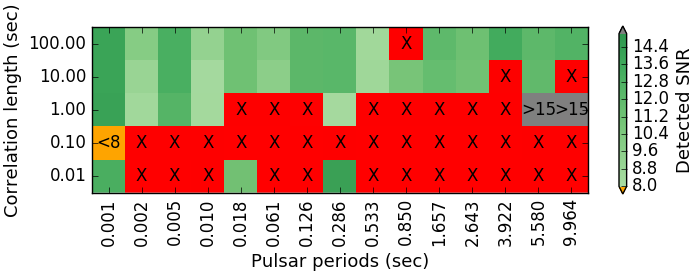}
	    \end{minipage}
    & 
	    \begin{minipage}{.4\textwidth}
	      \centering	    
	      \includegraphics[height=2.00cm]{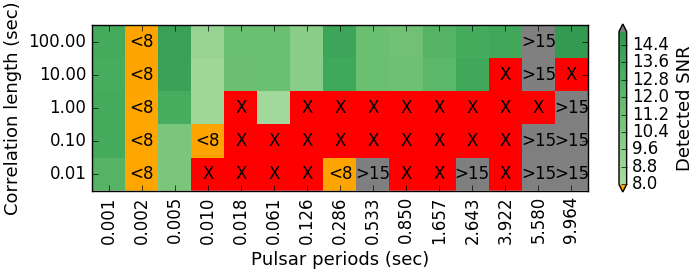}
	    \end{minipage}
    \\ \hline
    
  \end{tabular}
  \captionof{figure}{The SNR at which pulsars were detected after files containing them (see experiments 5 and 6 in Table~\ref{tb:Experiments}) were processed by four additional pipelines in \texttt{SIGPROC}.  The red squares represent missed pulsars, the orange squares represent detected pulsars with SNRs below the default threshold level of 8 and the grey squares represent detected pulsars with SNRs above the average maximum SNR of 15.}\label{tbl:SensitivitySIGPROC2}
\end{table*}

\section{Discussion}
With the advent of instruments like the Square Kilometre Array, real-time processing will become essential. Therefore, it is crucial that the pipeline employed to do this processing is optimal from the start. The purpose of the analysis in this paper was to investigate what improvements to current pulsar search pipelines are necessary before embarking on the development of a new real-time processing pipeline that is adept at dealing with the demands posed by this new era of pulsar astronomy.

This analysis demonstrated that non-stationary Gaussian noise processes with different correlation lengths lead to an increase in the number of false detections per true pulsar detection because of the static threshold applied in the power spectrum to distinguish between possible pulsar candidates and noise, i.e. non-stationary Gaussian noise is partly to blame for the so called 'crisis' in candidate selection \citep{lyon2016fifty}. In order to reduce the high number of false positives, \texttt{SIGPROC} as well as \texttt{PRESTO} employ spectrum whitening methods. Our analysis has revealed how these methods decrease the number of false positives per true positive at the cost of a loss in sensitivity and detection significance to long-period pulsars. 

The spectrum whitening techniques assessed in this analysis suppress the power in the lower frequencies to conform to the power levels of the higher frequencies. Consequently, the spectral power of real signals from slowly rotating pulsars is attenuated along with the noise. This analysis serves as evidence that there is room for improvement in the effectiveness of the current spectrum whitening methods. Instead of forcing the spectrum to be uniform in the lower frequencies, the solution should rather be to accurately model the noise both in the spectral and in the time domain. In fact we have shown that applying a 10~s moving average filter in the time domain resulted in a greater number of detections of long period pulsars. Consequently, leveraging moving averages on streaming data can help meet future real-time processing requirements whilst increasing surveys' sensitivity to long period pulsars.    

As we discussed earlier, the results presented in Figure~\ref{fig:Question6}, Figure~\ref{tbl:SensitivityPresto}, Figure~\ref{tbl:SensitivitySIGPROC1} and Figure~\ref{tbl:SensitivitySIGPROC2}, are based on single realizations of period-length scale combinations. However, we have sampled the standard deviation of the significance of these detections for several cases, and we conclude that although the picture may change for different realizations and different initial S/N values of the injected pulsars, the areas in the plots which are most affected remain the same.

In this analysis we dedispersed all the files at the same DM as what we injected the pulsars at (DM = 68 pc cm$^{-3}$). However, a subset of the files we dedispersed at four additional DM values, namely 0, 20, 150 and 300 pc cm$^{-3}$. Dedispersing the files at these four additional DMs allowed us to confirm that the number of false positives detected  by the pulsar search pipelines for the files containing both non-stationary noise and RFI are greatest when the filterbank files are not dedispersed and decreases as one moves away from 0 DM. However, the number of false positives detected by the pulsar search pipelines is very similar for the five DMs used to dedisperse the data. Consequently, the number of false positives detected for files containing only non-stationary noise is similar irrespective of the DM used to dedisperse the data.

In this analysis it was demonstrated that the RFI detection algorithm in \texttt{PRESTO} is very sensitive to the interplay between integration length over which the statistics of the filterbank files are computed and the rejection thresholds both in time and frequency of said statistics. For off-line processing this interplay can be fine tuned so that most RFI at different brightness levels can be detected and masked. However, for the real-time detection of RFI this exploration of parameter space is not always possible because of the time constraint as well as the dynamic nature of the RFI environment. 

There are a multitude of modules each placed strategically throughout current pulsar search pipelines for detecting different sources of RFI. Most of these RFI detection algorithms are largely amplitude-based and are therefore very sensitive to non-stationary baselines. Consequently, data which contain no RFI but which have higher than average mean and standard deviation are flagged as RFI. This analysis demonstrated that by flagging and replacing blocks of non-stationary data which contain no RFI or weak RFI may result in short period pulsars being attenuated below the detection threshold. Hence, there is a need for algorithms that can simultaneously normalise a non-stationary baseline and excise RFI signals superposed on said baseline without compromising the data that is not affected.


\section{Conclusion}
\label{sec:Conclusion}
This paper gives a unified view of a typical pulsar search system. Moreover, it delves into the particulars of the algorithms available in the pulsar search software packages \texttt{SIGPROC} and \texttt{PRESTO} for spectrum whitening.


This analysis accords with the \cite{b31} PALFA sensitivity analysis that non-stationary noise and weak RFI leads to an increase in the number of false positives and lower sensitivity for long period pulsars. These two effects have resulted in overestimates of survey production. 

The severe degradation of the detection significance is partly due to frequency dependent noise and partly due ot the attenuating nature of the spectrum whitening algorithms implemented in pulsar search software. Both these effects serve as explanation for why so many detectable long period normal pulsars are missed by pulsar search pipelines.

The analysis revealed that an increase in sensitivity was achieved when the data were de-trended with a moving average filter with a window size larger than the slowest pulsar. However, it should be noted that the efficacy of the filter is dependent on the filter window size relative to the correlation length of the non-stationary noise process. 

Following from the results of this paper it is now feasible to investigate methods for normalising a varying baseline as well as addressing the question of how to decouple the red noise from the signal without attenuating the detection significance. The effectiveness of these methods can be determined by applying them to the files created for this sensitivity analysis and then re-processing the normalised and modified files with the same pulsar search pipelines.  
\section*{Acknowledgements}
The authors wish to thank Paul Brook, Marisa Geyer, Jayanth Chennamangalam and Chris Williams for useful discussions.  Additionally, the authors would like to thank Sean Passmoor, Lindsay Magnus and Justin Jonas at SKA South Africa for supplying the necessary RFI information required for this analysis.  E. van Heerden would like to thank Bernard van Heerden for editorial revisions and acknowledges with gratitude the Commonwealth Scholarship Commission in the UK for providing financial support for this work. Finally, the authors would like to thank Scott Ransom, for useful comments that have helped improve the manuscript significantly.




\bibliographystyle{mnras}
\bibliography{AFrameworkforAssessingthePerformanceofPulsarSearchPipelines.bib} 

~\

%
%


\bsp	
\label{lastpage}
\end{document}